\documentclass[manuscript,screen,review=false,nonacm]{acmart}

\usepackage[normalem]{ulem}

\usepackage[ruled,vlined,linesnumbered]{algorithm2e}
\usepackage{amsmath}
\usepackage{amsthm}
\usepackage{array}
\usepackage{colortbl}
\usepackage{listings}
\usepackage{multirow}
\usepackage{graphicx}
\usepackage{braket}

\newtheorem{proposition}{Proposition}

\AtBeginDocument{%
  }

\setcopyright{acmcopyright}
\copyrightyear{2018}
\acmYear{2018}
\acmDOI{XXXXXXX.XXXXXXX}

\acmConference[Conference acronym 'XX]{Make sure to enter the correct
  conference title from your rights confirmation emai}{June 03--05,
  2018}{Woodstock, NY}
\acmPrice{15.00}
\acmISBN{978-1-4503-XXXX-X/18/06}

\begin{document}

\title{Quantum Advantage for All}

\author{Christoph M.~Kirsch}
\affiliation{
  \institution{University of Salzburg} 
  \country{Austria}                    
}
\affiliation{
  \institution{Czech Technical University, Prague} 
  \country{Czech Republic}             
}
\email{ck@cs.uni-salzburg.at}          

\author{Stefanie Muroya Lei}
\affiliation{
  \institution{Czech Technical University, Prague} 
  \country{Czech Republic}             
}
\email{stefanie.muroya@ucsp.edu.pe}


\begin{abstract}
We show that the algorithmic complexity of any classical algorithm written in a Turing-complete programming language polynomially bounds the number of quantum bits that are required to run and even symbolically execute the algorithm on a quantum computer. In particular, we show that any classical algorithm $A$ that runs in $\mathcal{O}(f(n))$ time and $\mathcal{O}(g(n))$ space requires no more than $\mathcal{O}(f(n)\cdot g(n))$ quantum bits to execute, even symbolically, on a quantum computer. With $\mathcal{O}(1)\leq\mathcal{O}(g(n))\leq\mathcal{O}(f(n))$ for all $n$, the quantum bits required to execute $A$ may therefore not exceed $\mathcal{O}(f(n)^2)$ and may come down to $\mathcal{O}(f(n))$ if memory consumption by $A$ is bounded by a constant. Our construction works by encoding symbolic execution of machine code in a finite state machine over the satisfiability-modulo-theory (SMT) of bitvectors, for modeling CPU registers, and arrays of bitvectors, for modeling main memory. The FSM is linear in the size of the code, independent of execution time and space, and represents the reachable machine states for any given input. The FSM may be explored by bounded model checkers using SMT and SAT solvers as backend. However, for the purpose of this paper, we focus on quantum computing by unrolling and bit-blasting the FSM into (1)~satisfiability-preserving quadratic unconstrained binary optimization (QUBO) models targeting adiabatic forms of quantum computing such as quantum annealing, and (2)~semantics-preserving quantum circuits (QCs) targeting gate-model quantum computers. Given a bound on $n$, both QUBOs and QCs are generated in $\mathcal{O}(f(n)\cdot g(n))$ time and space. From that we derive two major insights: (1)~through our QUBOs and in particular their relatively compact size, real quantum annealers can now execute simple but real code even symbolically, yet only with potential but no guarantee for exponential speedup, and (2)~through our QCs used as oracles, Grover's algorithm applies to symbolic execution of arbitrary code, guaranteeing at least in theory a quadratic speedup of symbolic execution on gate-model quantum computers.
\end{abstract}

\maketitle

\section{Introduction}

We have designed and implemented \emph{unicorn}\footnote{\label{fot:unicorn}Original prototype in C and Python (discontinued): \url{https://github.com/cksystemsteaching/selfie/tree/unicorn}}\footnote{Standalone prototype in Rust (in development as of 2022): \url{https://github.com/cksystemsgroup/unicorn}}, an open-source toolchain that enables symbolic execution of machine code through bounded model checking as well as adiabatic and gate-model quantum computing. Unicorn consists of three components: the \emph{BEATOR} frontend generating finite state machines for symbolic execution through bounded model checking (Section~\ref{sec:classical}) as well as the \emph{QUBOT} backend generating quadratic unconstrained binary optimization (QUBO) models for adiabatic quantum computing (Section~\ref{sec:quantum}) and the \emph{QUARC} backend generating quantum circuits (QCs) for gate-model quantum computing (Section~\ref{sec:quarc}).

Given a program $P$ written in a Turing-complete subset of C, BEATOR encodes symbolic execution of 64-bit and 32-bit RISC-V machine code generated from $P$ in a finite state machine over the theory of bitvectors for modeling CPU registers and arrays of bitvectors for modeling main memory. The FSM represents the reachable machine states for any given input. Each transition in the FSM corresponds to executing one machine instruction (or reading one machine word as input). Which instruction is determined by a Boolean flag called pc flag, one for each instruction in the code. Without any modifications to the code, BEATOR scans the code in a linear fashion two times, first generating combinational circuits for register and memory updates (data flow), and then generating combinational circuits for pc flag updates (control flow).

BEATOR is fast, generating the FSM in $\Theta(|P|)$ time and space. Optionally, given a bound $m$ on memory size, BEATOR may generate an equivalent FSM over just the theory of bitvectors without arrays of bitvectors in $\Theta(m\cdot|P|)$ time and space by unfolding all main memory access into pure register access. For this purpose, BEATOR generates upon each memory access a combinational circuit that switches over all $m$ memory addresses to return the correct memory entry. BEATOR outputs FSMs in a file format called BTOR2~\cite{BTOR2} for which a bounded model checker called \texttt{btormc}~\cite{BTOR2} exists that was instrumental in debugging and validating BEATOR.

Given a machine state $M$ and a bound $n$ on execution steps, QUBOT unrolls the FSM generated by BEATOR $n$ times, propagates constants, and bit-blasts the result into a combinational circuit that is encoded in a satisfiability-preserving QUBO model $B$ such that any solution of $B$ contains input on which $P$ runs into $M$ executing no more than $n$ instructions, and conversely, if $B$ has no solution, there is no input on which $P$ runs into $M$ executing no more than $n$ instructions.

QUBOT is also fast, generating the model in $\mathcal{O}(n\cdot(n\cdot|P|))$ time and space where $n$ also bounds memory size since each machine instruction may only expand the machine state by a constant amount of bits. Moreover, if memory consumption of $P$ is bounded by a constant, QUBOT's time and space complexity comes down to $\mathcal{O}(n\cdot|P|)$. QUBOT also supports unfolding main memory access. Doing that in QUBOT rather than BEATOR may in practice result in QUBO models with fewer quantum bits. QUBOT stores QUBO models in our custom-designed format for which we developed an auxiliary tool for debugging and validating QUBOT by evaluating QUBO models on known inputs. We also implemented support in QUBOT for deploying QUBO models on real quantum annealers from D-Wave systems. Section~\ref{sec:exp} has the details.

QUARC does essentially the same as QUBOT, even with slightly better time and space complexity if considering machine word size, but generates a semantics-preserving quantum circuit $Q$, rather than a QUBO model, such that $Q$ outputs 1 for input on which $P$ runs into $M$ executing no more than $n$ instructions, and 0 otherwise. Unlike QUBOT, QUARC relies on BEATOR for unfolding main memory access. QUARC outputs OpenQASM~\cite{qasm}, an open standard for specifying quantum circuits with tool support for validation, simulation, and deployment on real gate-model quantum computers.

Unicorn improves the state of the art in at least three dimensions with the following contributions:

\begin{enumerate}
\item (Software) Unicorn is the first tool that translates software written in a Turing-complete programming language that supports unbounded dynamic stack and heap allocation (procedures, malloc). Unicorn targets both adiabatic and gate-model quantum computing. The state of the art in that dimension is an algorithm that translates a subset of C that resembles a hardware description language that is not Turing-complete and does not support dynamic memory allocation~\cite{ASPLOS19}. Moroever, that algorithm does not support symbolic execution and bounded model checking and only targets quantum annealers.

\item (Space) Unicorn generates the to-date asymptotically most compact models, which are smaller by a linear factor in execution steps over the state of the art which is, to the best of our knowledge, a tool called CheckFence~\cite{CheckFence} for finding concurrency bugs. CheckFence transforms code into register SSA form prior to translation into models related to our FSMs. Unicorn avoids register SSA form by generating circuits that switch over all machine instructions that update a given register.

\item (Time) Unicorn may eventually speed up symbolic execution (1)~in practice by generating compact QUBO models that can already be solved by existing quantum annealers, and (2)~in theory by targeting gate-model quantum computing and thus guaranteeing quadratic speedup in finding inputs through Grover's algorithm~\cite{Grover96}.
\end{enumerate}

Unicorn's improvements over existing work~\cite{ASPLOS19,CheckFence} are not achieved by mere extension or refinement. Instead, unicorn and in particular BEATOR are based on a novel, uniform encoding of control and data flow as well as dynamic memory access developed from scratch which suppports symbolic execution of arbitrary code through bounded model checking, and quantum computing using QUBOT and QUARC as backend.

While all of unicorn is available as open source, its complexity is significant making both its presentation and proof of correctness difficult. The latter remains future work that is likely going to involve proof assistants and automation. For the presentation of unicorn we focus on algorithmic complexity of all relevant components and explain algorithmic details by example using the same running example throughout the paper. Detailed pseudo code of representative versions of BEATOR, QUBOT, and QUARC can be found in supplementary material.

\subsection*{Quantum Programming Models}

The background on quantum computing necessary to follow this paper requires understanding the logics but not the physics of the quantum programming models that are relevant here. As previously mentioned, we distinguish adiabatic quantum computing, in particular quantum annealers from gate-model quantum computers. Quantum annealers tend to have at least one order of magnitude more quantum bits (qubits) than existing gate-model machines, and are therefore an interesting, working target for unicorn using QUBOT as backend. In Section~\ref{sec:exp}, we report on our running example actually executing symbolically on a real quantum annealer by D-Wave systems. Gate-model machines are also an interesting target for unicorn with QUARC as backend but only for obtaining theoretical results, at least for now. Unfortunately, we have no access to a real gate-model machine.

The programming models of quantum annealers and gate-model machines that we use here are QUBO models and quantum circuits in OpenQASM~\cite{qasm}, respectively. A QUBO model is a binary quadratic function for which we would like to find a solution, that is, an assignment of its binary variables such that the function evaluates to zero, or else know that there is no solution. The problem of solving QUBO models is NP-hard. A quantum annealer, on the other hand, features a number of qubits where each qubit has a programmable bias towards 0 or 1. Moreover, each qubit can be programmed to be entangled with another qubit. Entanglement is a phenomenon in quantum mechanics that allow two qubits to correlate or anticorrelate their value. Both, bias and entanglement are programmable by normalized real values. A quantum annealer can solve a QUBO model by associating each binary variable $x$ of the model with a unique qubit $q$ on the annealer. Here, the terms (binary) variable and qubit can therefore be seen as logically synonymous. The linear factor of $x$ in the model is represented by the bias of $q$ on the annealer. A bi-linear\footnote{an example of a bi-linear factor is $-2xy$ whereas a quadratic factor such as $-2x^2$ is a linear factor if $x$ is a binary variable because then $-2x^2=-2x$} factor of $x$ with another variable $y$ in the model is represented by the entanglement of $q$ and the qubit associated with $y$ on the annealer.

Once linear and bi-linear factors are configured as bias and entanglement, respectively, quantum annealing refers to the process of moving from a high energy state in which all qubits are in superposition to a low (ground) energy state in which all qubits are 0 or 1 and represent a solution of the QUBO model with some probability that can be increased by repeating the process and tuning the model. The time to find out if there is a solution or not and, if there is, find a solution, depends on the model. In particular, the time is inversely proportional to the minimum gap of the model, which is the difference in energy between the ground state and the first excited state of the model~\cite{AQC}. In short, the closer non-solutions are to the ground state the longer it takes to find an actual solution. Another issue is that qubits can usually not be entangled with all other qubits. This can be addressed by representing a single binary variable with multiple physical qubits known as ancillae. Thus annealing QUBO models also involves solving an NP-hard problem called \emph{minor embedding} that searches for the least number of ancillae needed to represent a QUBO. However, minor embeddings can be solved efficiently and sufficiently accurate using heuristics. The following related work section and Section~\ref{sec:exp} have more on that.

Understanding quantum circuits and gate-model machines is more involved than understanding QUBO models and quantum annealers. However, since with QUARC we only translate combinational circuits to quantum circuits, we only need to know that any combinational circuit can be translated to a semantics-preserving quantum circuit of same asymptotic size. A gate-model machine can then be programmed to solve another NP-hard problem: finding inputs to a quantum circuit that make the circuit output, say, 1. Here, the time to do so may in the worst case be exponential in the number of input bits but, without any further information on the circuit, can be reduced at least by a quadratic factor using Grover's algorithm~\cite{Grover96}. Independently of that, the key challenge in QUARC as well as in QUBOT is to reduce the need for qubits so that the generated QUBO models and quantum circuits eventually fit on real quantum hardware.

\section{Related Work}
\label{sec:rel}

We found inspiration from two, separate lines of thought that we bring together here: (1)~translation of classical code to quantum annealers~\cite{ASPLOS19,HPEC19} and (2)~symbolic execution~\cite{SYMEX}, from systems such as KLEE~\cite{KLEE} and S2E~\cite{S2E}, and from bounded model checking~\cite{BMC}, such as the bounded model checker \texttt{btormc}~\cite{BTOR2}. Satisfiability Modulo Theory (SMT) solvers such as Z3~\cite{Z3} and boolector~\cite{Boolector}, and the fact that SAT and SMT formulae can be solved using quantum annealers have helped us as well~\cite{Bian17,Su16,King15,Weaver14,Bonet07}.

While BEATOR and QUBOT together logically resemble translation of Verilog to QMASM via EDIF~\cite{ASPLOS19}, there are a number of technical advancements as well as an important principled difference: firstly, unicorn translates a Turing-complete subset of C, in particular code using dynamic memory, and not Verilog~\cite{ASPLOS19} or domain-specific C code targeting quantum annealers~\cite{HPEC19}. Secondly, unlike~\cite{ASPLOS19}, it supports symbolic execution and bounded model checking not just on the theory of bitvectors but, most importantly, on the theory of arrays of bitvectors as well which is key to supporting dynamic memory allocation. Thirdly, it enables relating classical algorithmic complexity with spatial quantum complexity by lifting translation of classical code~\cite{ASPLOS19} to symbolic execution and bounded model checking on quantum computers. However, unicorn currently defers handling minor embeddings to a D-Wave library\footnote{\label{fot:dwave}\url{www.dwavesys.com}} at the expense of improved annealing performance~\cite{ASPLOS19}.

BEATOR improves asymptotic model size over CheckFence~\cite{CheckFence} by a linear factor making models generated by BEATOR linear in code size. Unlike CheckFence, BEATOR avoids transforming code into register SSA form and instead translates code directly to combinational circuits that switch over all instructions that update a given register. However, which choice provides better performance in practice is unclear.

BTOR2~\cite{BTOR2} extends SMT-LIB~\cite{SMTLIB} with sequential operators that inspired us in how to avoid the path and state explosion problems when encoding symbolic execution of arbitrary code. Translating BTOR2 to quadratic unconstrained binary optimization (QUBO) models takes us to the domain of QUBO problems and solvers. QUBO is NP-hard~\cite{Date2019} and has numerous applications beyond quantum annealing~\cite{Hao14}. QUBO problems can also be solved by algorithms that target gate-model quantum computers~\cite{Ehsan17,GroverQUBO,ImaginaryTimeEvol,VQA}, and other classical algorithms such as Steepest Gradient Descent~\cite{Meza10} and Tabu Search~\cite{Gendreau05} as well as variations of Tabu Search~\cite{Glover98,Zhou13,Gerald11,Misevicius05,Battiti97}, Simulated Quantum Annealing (SQA)~\cite{Hariyama20,Crosson16}, Global Equilibrium Search (GES)~\cite{Shylo12,Shylo10}. SQA algorithms use Markov Chain Monte Carlo methods~\cite{Crosson16} and classical accelerators such as GPUs and FPGAs~\cite{Hariyama20} to deal with the problem that current quantum computers feature only relatively small amounts of qubits. Optimization techniques found in QUBO solvers but also SMT solvers and bounded model checkers as well as in compilers may eventually help reducing the number of qubits when translating RISC-V via BTOR2 to QUBO.

Moving on to quantum computing, there are essentially two mainstream models of quantum computers or Noisy Intermediate-Scale Quantum (NISQ) devices: quantum annealing (QA) and gate-model (GM) machines. Our results directly apply to both, QA using QUBOT and GM using QUARC, while both are also at least in principle interchangable~\cite{Wilson21,Atrey19}. In the NISQ-devices market~\cite{Crosson21}, QA greatly outperforms GM in number of available qubits with the most recently released Quantum Processor Unit (QPU) featuring 5k qubits, and the announcement of a QPU with 7k qubits. Moreover, QA has recently been shown to return global optima with higher probability than GM while both, QA and GM, do not provide fairness if multiple ground states exist~\cite{Pelofske2021}. QA has already been applied in various domains other than executing arbitrary code~\cite{Tabi20,Wang20,Hayato17,Adachi15,Andrew14}.

In principle, QA can solve any QUBO model. However, for QA to work QUBO models still need to be minor-embedded onto the particular architecture of a machine where not all qubits can be entangled with all other qubits. Finding proper minor embeddings~\cite{Gideon20,Boothby20,Zbinden20,Date2019} is an NP-hard problem itself and important to enable quantum annealing and increase performance~\cite{ASPLOS19}. Improving the QUBO models to enhance the solution space~\cite{Pakin21,Lewis17,Tanburn15,Bian14}, considering how various parameters in the annealing process affect results~\cite{Grant21,King14}, and addressing the problem of too many binary variables versus too few available qubits~\cite{Kurowski20} is important as well.

\section{Classical Modeling with BEATOR}
\label{sec:classical}

\begin{figure}
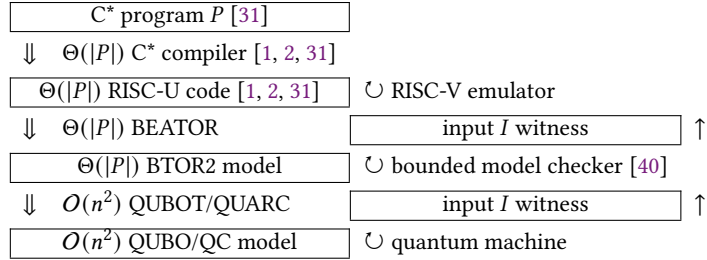

\centering
\begin{tabular}{rllr}
\cline{1-2}
\multicolumn{2}{|c|}{C* program $P$~\cite{Onward17}} & \\
\cline{1-2}
\vspace*{-2ex}\\
$\Downarrow$ & $\Theta(|P|)$ C* compiler~\cite{ASE21,MoreVMs18,Onward17} \\
\vspace*{-2ex}\\
\cline{1-2}
\multicolumn{2}{|c|}{$\Theta(|P|)$ RISC-U code~\cite{ASE21,MoreVMs18,Onward17}} & $\circlearrowright$ RISC-V emulator \\
\cline{1-2}
\vspace*{-2ex}\\
\cline{3-3}
$\Downarrow$ & $\Theta(|P|)$ BEATOR &
\multicolumn{1}{|c|}{input $I$ witness} & $\uparrow$ \\
\cline{3-3}
\vspace*{-2ex}\\
\cline{1-2}
\multicolumn{2}{|c|}{$\Theta(|P|)$ BTOR2 model} & $\circlearrowright$ bounded model checker~\cite{BTOR2} \\
\cline{1-2}
\vspace*{-2ex}\\
\cline{3-3}
$\Downarrow$ & $\mathcal{O}(n^2)$ QUBOT/QUARC &
\multicolumn{1}{|c|}{input $I$ witness} & $\uparrow$ \\
\cline{3-3}
\vspace*{-2ex}\\
\cline{1-2}
\multicolumn{2}{|c|}{$\mathcal{O}(n^2)$ QUBO/QC model} & $\circlearrowright$ quantum machine \\
\cline{1-2}
\end{tabular}
\caption{Unicorn toolchain and workflow}
\label{tbl:workflow}
\end{figure}

Fig.~\ref{tbl:workflow} provides an overview of the unicorn toolchain. We first introduce, by running example, the part of the toolchain that translates a Turing-complete subset of C called C*~\cite{Onward17} via a subset of RISC-V~\cite{RISCV} called RISC-U~\cite{ASE21} to a subset of BTOR2~\cite{BTOR2}. The compiler from C* to RISC-U has been implemented in C* independently of unicorn and is used here as is~\cite{ASE21,MoreVMs18,Onward17}. BEATOR, the RISC-U-to-BTOR2 translator we introduce here, is written in C* in around 3.5-KLOC utilizing the compiler as frontend. Program inputs found during symbolic execution can be validated by any RISC-V emulator including a RISC-U emulator that is part of the compiler~\cite{ASE21,MoreVMs18,Onward17}. BTOR2 models can be validated by the bounded model checker \texttt{btormc}~\cite{BTOR2}.

\subsection{C*~\cite{Onward17}}

\begin{figure}
\centering
\begin{lstlisting}[frame=single,language=C]
uint64_t* x;
uint64_t main() { uint64_t a;
  x = malloc(1); // rounded up to 8
  // touch to trigger page fault here
  *x = 0;
  // read 1 byte from console into x
  read(0, x, 1);
  // copy from heap to stack segment
  a = *x;
  // decrement input until <= '0'
  while (a > '0')
    a = a - 1;
  // segmentation fault on input '1'
  if (a == *x - 1) // '0' == '1' - 1
    // segfault: '0' != 0
    a = *(x + a);
  return 0;
}
\end{lstlisting}
\caption{Running example of a C* program}
\label{fig:source}
\end{figure}

Consider the source code of the running example in Fig.~\ref{fig:source}. The program reads a single byte from the console keyboard and returns zero as exit code unless the user presses \texttt{1} in which case a segmentation fault may be triggered by the attempt to read from unallocated memory with \texttt{*(x + a)} since \texttt{a} is then \texttt{'0'} which is \texttt{48}, and not \texttt{0}. Below we consider certain types of unsafe memory access triggering segmentation faults as well as non-zero exit codes and division by zero as machine states to look for. However, the toolchain is able to compute program input that leads to any given machine state. Program input is all \texttt{read} input. Once some input has been determined the code can be executed on that input to validate if a machine state is actually reached using, for example, the RISC-U emulator.

The code is written in C* which is a tiny subset of C that has originally been developed for educational purposes~\cite{ASE21,MoreVMs18,Onward17}. The example essentially features all elements of C* except procedure calls. C* only features two data types, \texttt{uint64\_t} and \texttt{uint64\_t*}, and five statements: assignment, \texttt{while}, \texttt{if}, \texttt{return}, and procedure call. There are the usual arithmetic and comparison operators but only for unsigned 64-bit integer arithmetic. Notably, there is only the unary dereference operator \texttt{*} to access heap memory. There are no arrays and no structs hence the name C*. Furthermore, C* supports integer, character, and string literals as well as global and local variables and procedure parameters. Lastly, there is \texttt{printf} library support and a total of five builtin procedures: \texttt{exit}, \texttt{open}, \texttt{read}, \texttt{write}, and \texttt{malloc}. Turns out that, because of its overall simplicity and in particular its focus on unsigned integer arithmetic, C* is well-suited as target for researching and prototyping tools for symbolic execution~\cite{SYMEX,KLEE,Z3,BTOR2}.

The part of the toolchain that is relevant here is written in C* and consists of a non-optimizing linear-time (modulo global (local) symbol table hash collisions (search)) C* compiler that targets RISC-U, a tiny subset of 64-bit RISC-V~\cite{RISCV}, as well as a RISC-U emulator and BEATOR which translates RISC-U code to BTOR2~\cite{BTOR2}. RISC-U binaries generated by the C* compiler are in ELF format and run not only on the RISC-U emulator but also on QEMU and actual 64-bit RISC-V hardware. The toolchain also fully supports 32-bit machines and generates 32-bit code by bootstrapping \texttt{uint64\_t} to \texttt{uint32\_t} and carefully avoiding integer overflows beyond 32 bits. Support of 32-bit binaries highlights the impact of machine word size on number of qubits in Section~\ref{sec:exp}. Thus everything below also applies to 32-bit machines and code.

\subsection{RISC-U ISA~\cite{ASE21,MoreVMs18,Onward17}}

\begin{figure}
\centering
\begin{lstlisting}[frame=single,morekeywords={ld,addi,sltu,beq,sub,sd,jal,mul,add}]
WHILE:ld t0,-8(s0)      //t0=a
      addi t1,zero,48   //t1='0'
      sltu t0,t1,t0     //t0='0'<a
      beq t0,zero,6[IF] //goto IF '0'=>a
      ld t0,-8(s0)       //t0=a
      addi t1,zero,1     //t1=1
      sub t0,t0,t1       //t0=a-1
      sd t0,-8(s0)       //a=a-1
      jal zero,-8[WHILE]//goto WHILE
IF:   ld t0,-8(s0)      //t0=a
      ld t1,-16(gp)     //t1=x
      ld t1,0(t1)       //t1=*x
      addi t2,zero,1    //t2=1
      sub t1,t1,t2      //t1=*x-1
      sub t0,t1,t0      //t0=*x-1-a
      addi t1,zero,1    //t1=1
      sltu t0,t0,t1     //t0=*x-1-a<1
      beq t0,zero,8[R0]//goto R0 *x-1!=a
A0:   ld t0,-16(gp)     //t0=x
      ld t1,-8(s0)      //t1=a
      addi t2,zero,8    //t2=8 (bytes)
      mul t1,t1,t2      //t1=a*8
A1:   add t0,t0,t1      //t0=x+a*8
SEGFL:ld t0,0(t0)       //t0=*(x+a)
      sd t0,-8(s0)      //a=*(x+a)
R0:   addi t0,zero,0    //t0=0
      addi a0,t0,0      //a0=0
      jal zero,1[EXIT]  //return 0 (a0)
\end{lstlisting}
\caption{RISC-U assembly fragment generated for the running example in Fig.~\ref{fig:source}}
\label{fig:assembly}
\end{figure}

Consider the RISC-U assembly fragment in Fig.~\ref{fig:assembly} which has been generated by the C* compiler for the \texttt{while} loop as well as the \texttt{if} and the \texttt{return} statements in the running example. The assembly fragment features 9 out of the 14 RISC-U instructions. Overall code size is linear in the size of the C* source. Labels and comments are provided manually.

\begin{table}
\centering
\begin{tabular}{|l|m{0.59\linewidth}|}
\hline
\multicolumn{2}{|c|}{Initialization} \\
\hline
  \texttt{lui rd,imm} & $rd \leftarrow imm * 2^{12}$; $pc \leftarrow pc + 4$ \\
  \cline{2-2}
  \texttt{addi rd,rs1,imm} & $rd \leftarrow rs1 + imm$; $pc \leftarrow pc + 4$ \\
\hline
\multicolumn{2}{|c|}{Memory} \\
\hline
  \texttt{ld rd,imm(rs1)} & $rd \leftarrow mem[rs1 + imm]$; $pc \leftarrow pc + 4$ \\
  \cline{2-2}
  \texttt{sd rs2,imm(rs1)} & $mem[rs1 + imm] \leftarrow rs2$; $pc \leftarrow pc + 4$ \\
\hline
\multicolumn{2}{|c|}{Arithmetic} \\
\hline
  \texttt{add rd,rs1,rs2} & $rd \leftarrow rs1 + rs2$; $pc \leftarrow pc + 4$ \\
  \cline{2-2}
  \texttt{sub rd,rs1,rs2} & $rd \leftarrow rs1 - rs2$; $pc \leftarrow pc + 4$ \\
  \cline{2-2}
  \texttt{mul rd,rs1,rs2} & $rd \leftarrow rs1 * rs2$; $pc \leftarrow pc + 4$ \\
  \cline{2-2}
  \texttt{divu rd,rs1,rs2} & $rd \leftarrow rs1 \mathrel{/}_{unsigned} rs2$; $pc \leftarrow pc + 4$ \\
  \cline{2-2}
  \texttt{remu rd,rs1,rs2} & $rd \leftarrow rs1 \mathrel{\%}_{unsigned} rs2$; $pc \leftarrow pc + 4$ \\
\hline
\multicolumn{2}{|c|}{Comparison} \\
\hline
  \texttt{sltu rd,rs1,rs2} &
    $rd \leftarrow \begin{cases}
        1, & \text{if}\ rs1 <_{unsigned} rs2 \\
        0, & \text{otherwise}
    \end{cases}$ \newline
    $pc \leftarrow pc + 4$ \\
\hline
\multicolumn{2}{|c|}{Control} \\
\hline
  \texttt{beq rs1,rs2,imm} &
    $pc \leftarrow \begin{cases}
        pc + imm, & \text{if}\ rs1 == rs2 \\
        pc + 4, & \text{otherwise}
    \end{cases}$ \\
  \cline{2-2}
  \texttt{jal rd,imm} &
    $rd \leftarrow pc + 4$; $pc \leftarrow pc + imm$ \\
  \cline{2-2}
  \texttt{jalr rd,imm(rs1)} &
    $tmp \leftarrow ((rs1 + imm) /_{unsigned} 2) * 2$; \newline
    $rd \leftarrow pc + 4$; $pc \leftarrow tmp$ \\
\hline
\multicolumn{2}{|c|}{System} \\
\hline
  \texttt{ecall} &
    system call number in $a7$, parameters in $a0$--$a3$, return value in $a0$ \\
\hline
\end{tabular}
\caption{Table of the 14 RISC-U instructions~\cite{Onward17}}
\label{tbl:riscu}
\end{table}

In order to familiarize yourself with the instruction set, consider Table~\ref{tbl:riscu} which shows syntax and semantics of the 14 RISC-U instructions. Similar to C*, RISC-U only supports unsigned 64-bit integer arithmetic, hence the name RISC-U.

A RISC-U machine has a 64-bit program counter denoted \texttt{pc}, 32 general-purpose 64-bit registers numbered \texttt{0} to \texttt{31} and denoted \texttt{zero}, \texttt{ra}, \texttt{sp}, \texttt{gp}, \texttt{tp}, \texttt{t0-t2}, \texttt{s0-s1}, \texttt{a0-a7}, \texttt{s2-s11}, \texttt{t3-t6}, and 4GB of byte-addressed main memory. Register \texttt{zero} always contains the constant value 0. RISC-U binaries only use up to 18 of the 32 general-purpose registers, namely \texttt{zero}, \texttt{ra} which stands for \emph{return address}, \texttt{sp} for \emph{stack pointer}, \texttt{gp} for \emph{global pointer}, \texttt{t0-t6} where the \texttt{t} stands for \emph{temporary}, \texttt{s0} where the \texttt{s} stands for \emph{saved}, and \texttt{a0-a3} and \texttt{a6-a7} where \texttt{a} stands for \emph{argument}.

RISC-U instructions are encoded in 32 bits (4 bytes) each and stored next to each other in memory such that there are two instructions per 64-bit double word. Memory, however, can only be accessed at 64-bit double-word granularity. The \texttt{d} in \texttt{ld} and \texttt{sd} stands for double word. Below we nevertheless refer to double word by machine or memory word or just word for brevity. The parameters \texttt{rd}, \texttt{rs1}, and \texttt{rs2} denote any of the general-purpose registers. The parameter \texttt{imm} denotes an \emph{immediate} value which is a signed integer in two's complement represented by a fixed number of bits depending on the instruction. The five builtin procedures of C* are implemented by five system calls that follow the ABI of Linux and the official RISC-V toolchain enabling RISC-U binaries to run on both platforms as well.

\begin{figure}
\centering
\begin{tabular}{r|c|c|l}
\cline{2-3}
& RISC-U segments & C* variables & \\
\cline{2-3}
\cline{2-3}
\texttt{0xFFFFFFC0}\_ & stack & & \_\texttt{s0} \\
\texttt{0xFFFFFFB8}\_ & $\downarrow$ & \texttt{a} & \_\texttt{s0-8}, \_\texttt{sp} \\
\cline{2-3}
& \multicolumn{2}{c|}{} & \\
& \multicolumn{2}{c|}{$\times$} & \\
\texttt{0x12008}\_ & \multicolumn{2}{c|}{} & \texttt{\_bump} \\
\cline{2-3}
& $\uparrow$ & & \\
\texttt{0x12000}\_ & heap & \texttt{*x} & \\
\cline{2-3}
\texttt{0x11010}\_ & \multicolumn{2}{c|}{$\times$} & \_\texttt{gp} \\
\cline{2-3}
& & \texttt{\_bump} & \_\texttt{gp-8} \\
\texttt{0x11000}\_ & data & \texttt{x} & \_\texttt{gp-16} \\
\cline{2-3}
\texttt{0x10228}\_ & \multicolumn{2}{c|}{$\times$} & \\
\cline{2-3}
\texttt{0x10000}\_ & code & & \\
\cline{2-3}
& \multicolumn{2}{c|}{$\times$} & \\
\cline{2-3}
\end{tabular}
\caption{4KB-page-aligned RISC-U memory segments and machine state when executing the RISC-U code in Fig.~\ref{fig:assembly}}
\label{fig:memory}
\end{figure}

The RISC-U memory layout is shown in Fig.~\ref{fig:memory}. Code starts by RISC-V convention at \texttt{0x10000} and ends for our running example at \texttt{0x10228}. The data segment starts 4KB-page-aligned after the code segment at \texttt{0x11000} and ends at \texttt{0x11010}, again for our running example. The data segment contains values of global variables, string literals, and big integer literals that require more than 32 bits up to 64 bits. The layout of the data segment does not change during code execution. Entries are addressed relative to register \texttt{gp} which is initialized by the executed code, here to \texttt{0x11010}, and then never changed. In our running example, the data segment contains two 64-bit memory words, one for the global variable \texttt{x} at \texttt{gp-16} and one for an internal variable \texttt{\_bump} at \texttt{gp-8} which facilitates a bump-pointer allocator for \texttt{malloc}. The \texttt{\_bump} pointer is initialized by the executed code to the start of the heap which is 4KB-page-aligned after the end of the data segment, here at \texttt{0x12000}. After executing \texttt{x = malloc(1);}, the \texttt{\_bump} pointer is increased by 8 bytes (64 bits), rounded up from 1 byte, to \texttt{0x12008}. The allocated address returned by \texttt{malloc} is the previous value of \texttt{\_bump} which is \texttt{0x12000}, hence making \texttt{x} point to the first 64-bit memory word \texttt{*x} on the heap. The byte read from the console keyboard by the \texttt{read(0, x, 1)} call is stored in the least significant byte of that memory word with the more significant bytes set to~0.

The stack segment starts at the address where register \texttt{sp} points to, here \texttt{0xFFFFFFB8}, which is the top of the call stack. The end of the stack segment is the highest address in main memory. Thus the call stack grows downwards to lower addresses while the heap grows upwards to higher addresses towards the stack. There is currently no \texttt{free} call in the system but it could be added without affecting modeling. The call stack and register \texttt{sp} are initialized by the boot loader. Local variables and procedure parameters on the call stack are accessed relative to register \texttt{s0} called the \emph{frame pointer}. Here, the local variable \texttt{a} is at the top of the stack at \texttt{s0-8} where \texttt{s0} is set to \texttt{0xFFFFFFC0} by the callee of the currently executed procedure, that is, here the \texttt{main} procedure.

In sum, the state of a RISC-U machine consists of the 64-bit program counter, the 18 general-purpose 64-bit registers, and the 64-bit memory words in the data, heap, and stack segments. BEATOR as well as QUBOT and QUARC must handle that state.

\subsection{BTOR2~\cite{BTOR2} (BEATOR2)}

BTOR2 is a formalism~\cite{BTOR2} for modeling finite state machines over the theory of bitvectors and arrays of bitvectors. Essentially, BTOR2 extends the bitvector fragment of SMT-LIB~\cite{SMTLIB} with sequential operators for states and state transitions. BTOR2 models are input to bounded model checkers~\cite{BMC} such as \texttt{btormc}~\cite{BTOR2}. Given a BTOR2 model, a bounded model checker determines whether there is input to the model such that it transitions from its initial state to a \texttt{bad} state in a given number of state transitions where \texttt{bad} is a BTOR2 operator that essentially specifies negated safety properties~\cite{BTOR2}.

\begin{table}
\centering
\begin{tabular}{|l|l|}
\hline
BEATOR2 ($\subset$ BTOR2) & Usage and Relation to RISC-U \\
\hline\hline
  \texttt{sort bitvec} $size$ &
    value types (64-bit machine word and 1-bit Boolean)\\
  \texttt{sort array} $sort$ $sort$ &
    memory type (64-bit elements, 64-bit address space) \\
\hline\hline
\multicolumn{2}{|l|}{Combinational Operators (related to SMT-LIB~\cite{SMTLIB})} \\
\hline\hline
  \texttt{constd} $sort$ $integer$ &
    initial values, static addresses, immediate values \\
\hline
  \texttt{add} $sort$ $x$ $y$ &
    models \texttt{add rd,$x$,$y$} \\
  \texttt{sub} $sort$ $x$ $y$ &
    models \texttt{sub rd,$x$,$y$} \\
  \texttt{mul} $sort$ $x$ $y$ &
    models \texttt{mul rd,$x$,$y$} \\
  \texttt{udiv} $sort$ $x$ $y$ &
    models  \texttt{divu rd,$x$,$y$} \\
  \texttt{urem} $sort$ $x$ $y$ &
    models  \texttt{remu rd,$x$,$y$} \\
  \texttt{ult} $sort$ $x$ $y$ &
    models \texttt{sltu rd,$x$,$y$} \\
  \texttt{uext} $sort$ $x$ $b$ &
    unsigned extension of \texttt{ult}, \texttt{input} to 64-bit words \\
  \texttt{slice} $sort$ $x$ $u$ $l$ &
    address down-scaling $u-l+1$ bits from bit $l$ to bit $u$ \\
\hline
  \texttt{ite} $sort$ $bool$ $x$ $y$ &
    \texttt{ite} cascades over pc flags for state updates \\
  \texttt{and} $sort$ $bool$ $bool$ &
    control flow of \texttt{beq} and \texttt{jalr} \\
  \texttt{not} $sort$ $bool$ & \texttt{jalr},
    $false$ branch of \texttt{beq} \\
  \texttt{eq} $sort$ $x$ $y$ & \texttt{jalr},
    $true$ branch of \texttt{beq} \\
\hline
  + \texttt{neq}, \texttt{ulte}, \texttt{ugt(e)} &
    bound checks, kernel model \\
\hline
  \texttt{read} $sort$ $mem$ $idx$ &
    models \texttt{ld rd,$idx$} \\
  \texttt{write} $sort$ $mem$ $idx$ $val$ &
    models \texttt{sd $val$,$idx$} \\
\hline\hline
\multicolumn{2}{|l|}{Sequential Operators (not in SMT-LIB~\cite{SMTLIB})} \\
\hline\hline
  \texttt{state} $sort$ $name$ &
    variable of $sort$ with $name$ \\
  \texttt{init} $sort$ $state$ $val$ &
    initial value $val$ of $state$ \\
  \texttt{next} $sort$ $state$ $val$ &
    transition of $state$ to $val$ \\
\hline
  \texttt{input} $sort$ &
    symbolic input variable used in \texttt{READ} syscall model \\
\hline
  \texttt{bad} $bool$ &
    $bool$ describes machine state \\
\hline
\end{tabular}
\caption{Table of all BEATOR2 declarations and combinational and sequential operators from BTOR2~\cite{BTOR2}}
\label{tbl:beator2}
\end{table}

The unicorn toolchain includes a RISC-U-to-BTOR2 translator called \emph{BEATOR} that generates models in a subset of BTOR2 called BEATOR2, as summarized in Table~\ref{tbl:beator2} including its usage and informal relation to RISC-U. Below, we nevertheless speak of BTOR2 rather than BEATOR2 despite the ambiguity. While BEATOR supports the \texttt{bad} operator, support of BTOR2 operators for specifying (global) fairness constraints and (negations of) liveness properties~\cite{BTOR2} is future work. BEATOR runs in time linear and generates models in space linear in the size of RISC-U code, and thus in the size of C* programs:

\begin{proposition}
\label{prop:beator}
Let $P$ be a C* program, $M$ a machine state, and $n$ a bound on number of executed machine instructions. Then, for all program input $I$, the RISC-U code $R$ generated from $P$ runs on $I$ into $M$ executing no more than $n$ machine instructions if and only if the \texttt{bad} state modeling $M$ in the BTOR2 model generated by BEATOR for $R$ is reachable on $I$ in no more than $n+\lfloor|I|/w\rfloor$ state transitions where $w$ is machine word size (reading input in RISC-U only takes a single \texttt{ecall} instruction whereas in the BTOR2 model reading input takes as many transitions as there are machine words in $I$).
\end{proposition}

Examples of machine states are return of non-zero exit codes, division by zero, and segmentation faults, that is, any, possibly input-dependent memory access outside of the data, heap, and stack segments. We added checking for segmentation faults to BEATOR to support QUBO modeling of memory access. Optionally, BEATOR can also generate additional checks such that memory accesses outside of memory blocks allocated by \texttt{malloc} are machine states to look for. However, those checks increase the size of the BTOR2 model by around~50\%.

The key idea of BEATOR is to separate control and data flow in RISC-U code as follows. For data flow, BEATOR generates zeroed 64-bit bitvectors, one for each general-purpose register, and an array of zeroed 64-bit bitvectors, one for each 64-bit memory word, indexed by a 64-bit bitvector as memory address. The array models RISC-U main memory where the data and stack segments are initialized exactly as a RISC-U bootloader initializes them prior to code execution. This also includes the stack pointer which is the only register that must be initialized by the bootloader. BEATOR supports a down-scaled linear address space and a segmenting MMU and RAM model that avoids arrays of bitvectors. QUBOT also models MMU and RAM which allows us to tradeoff translation complexity, as described in the next section and in Section~\ref{sec:exp}.

\begin{figure}
\centering
\begin{lstlisting}[frame=single,morekeywords={sort,bitvec,array,zero,state,init,constd,ite,add,next,beq,ld,addi}]
1 sort bitvec 1  ; Boolean
2 sort bitvec 64 ; 64-bit machine word
3 sort array 2 2 ; 64-bit physical memory
10 zero 1 ...
//... register states 200-231 ...
200 zero 2 zero // register $0 is always 0 ...
203 state 2 gp ; register $3 ...
205 state 2 t0 ; register $5
206 state 2 t1 ; register $6
//... program counter states ...
16603600 state 1      // beq t0,zero,8[R0]:
16603601 init 1 16603600 10
16604000 state 1   // A0:ld t0,-16(gp) ...
16606800 state 1   // R0:addi t0,zero,0
//... 64-bit memory (data,heap,stack):
20000000 state 3 physical-memory
  loading data,heap,stack into memory:
20000001 init 3 20000000 17380002
//... data flow ...   A0:ld t0,-16(gp):
36604000 constd 2 -16
36604001 add 2 203 36604000
36604003 read 2 20000000 36604001
36604004 ite 2 16604000 36604003 36603202
//...                 A1:add t0,t0,t1:
36605600 add 2 205 206
36605601 ite 2 16605600 36605600 36604004
//...              SEGFL:ld t0,0(t0):
36606002 ite 2 16606000 36606001 36605601
//...                 R0:addi t0,zero,0:
36606800 ite 2 16606800 200 36606002
//... updating registers ...
60000005 next 2 205 36606800 t0
\end{lstlisting}
\caption{Data-flow fragment of the BTOR2 model for the running example}
\label{fig:data}
\end{figure}

For control flow, BEATOR generates zeroed 1-bit bitvectors, called pc flags, one for each instruction in the RISC-U code, for modeling the program counter. The pc flag for the instruction at the entry point of the code is the only pc flag initialized to \texttt{1} indicating that this instruction is the first to execute. From then on, control flows through the model by resetting the pc flag of the current instruction, after executing it, and then setting the pc flag of the next instruction to execute. Thus the invariant here is that at all times all pc flags are \texttt{0} except one. The alternative to pc flags is to represent the program counter explicitly by a 64-bit bitvector (32 bits are actually enough here but require scaling) and then control execution of instructions by comparing that bitvector with their constant addresses in memory. The advantage is that we only need one fixed-size bitvector instead of as many pc flags as there are instructions. However, model size still remains linear in code size and model checking performance may be negatively affected too. Exploring that alternative, also in QUBOT, remains future work.

Consider Fig.~\ref{fig:data} which shows a data-flow fragment of the BTOR2 model generated from the running example in Fig.~\ref{fig:source}. In BTOR2, comments are single-line comments that begin with a semicolon. Comments that begin with a double slash are not BTOR2 format and only inserted by us by hand for documentation. Each line begins with a node (line) identifier (nid) which must be larger than any previous nid. After the nid there is a keyword that identifies a BTOR2 operator followed by its arguments which may only be nids and integer literals. Forward references to larger nids than the current nid are not allowed.

\begin{figure}
\centering
\begin{lstlisting}[frame=single,morekeywords={one,eq,not,next,and,ite,beq,ld,sd,addi}]
11 one 1
//... data flow ...
36603600 eq 1 205 200   // $t0==$zero
36603601 not 1 36603600 // $t0!=$zero
//... control flow ...
//             beq t0,zero,8[R0]:
56603600 next 1 16603600 16603200
//          A0:ld t0,-16(gp):
56604000 and 1 16603600 36603601
56604001 next 1 16604000 56604000
// ...         sd t0,-8(s0):
56606400 next 1 16606400 16606000
// ...      R0:addi t0,zero,0:
56606800 and 1 16603600 36603600
56606801 ite 1 56606800 11 16606400
56606802 next 1 16606800 56606801
\end{lstlisting}
\caption{Conditional control-flow fragment of the BTOR2 model for the running example}
\label{fig:control}
\end{figure}

The model begins with sort (type) declarations of pc flags and memory word bitvectors as well as the memory array in nids \texttt{1-3}. Sorts are applied strictly by \texttt{btormc} which turns out to be helpful for debugging BEATOR. The \texttt{2 2} in line \texttt{3 sort array 2 2} refer to nid \texttt{2} as the index and element sorts of the array. Line \texttt{10 zero 1} declares the constant \texttt{0} of sort 1-bit bitvector. Lines \texttt{200-231} declare \texttt{state} variables for the 32 registers \texttt{0-31}. Next are \texttt{zero}ed \texttt{state} declarations of the pc flags where each nid begins with digit \texttt{1}, followed by the address of the instruction in decimal at runtime, followed by \texttt{00} and \texttt{01} for declaration and \texttt{init}ialization, respectively. For example, the line at nid \texttt{16603600} declares the pc flag for \texttt{beq t0,zero,8[R0]} which is stored in memory at \texttt{0x101F4} or \texttt{66036} in decimal. Memory is declared in \texttt{20000000 state 3 physical-memory} and initialized in \texttt{20000001 init 3 20000000 17380002} where nid \texttt{17380002} refers to the initial state of memory (not shown).

\begin{figure}
\centering
\begin{lstlisting}[frame=single,morekeywords={zero,constd,sort,bitvec,input,uext,state,init,ite,eq,add,write,next,ugte,ult,and,bad}]
20 zero 2 ...
22 constd 2 2 ...
//... 1-byte input
71 sort bitvec 8 ; 1 byte ...
81 input 71      ; 1 byte ...
91 uext 2 81 56 // extending input to 64 bits
//... register states ...
202 state 2 sp ; register $2 ...
210 state 2 a0 ; register $10
211 state 2 a1 ; register $11
//... read system call ...
42000001 ite 2 42000000 211 36609200 ...
42000007 eq 1 42000006 22 // inc == 2
42000008 ite 2 42000007 92 91 ...
42000019 eq 1 42000006 28 // inc == 8
42000020 ite 2 42000019 98 42000018
42000021 add 2 211 210 ; $a1 + $a0
//  memory[$a1 + $a0] = input:
42000022 write 3 20000000 42000021 42000020
//... brk system call:
45000001 state 2 brk-bump-pointer
//... updating physical memory:
70000000 next 3 20000000 42000028
//... address >= current end of heap:
80000006 ugte 1 44000001 45000001
// address < current start of stack:
80000007 ult 1 44000001 202
80000008 and 1 80000006 80000007
// access between heap and stack:
80000009 bad 80000008 b2
\end{lstlisting}
\caption{System call and \texttt{bad} state fragment of the BTOR2 model for the running example}
\label{fig:system}
\end{figure}

Then, there are \texttt{ite} (if-then-else) cascades that encode per-instruction data flow where each nid begins with digit \texttt{3}, followed by an \texttt{ite} expression that either selects the data flow of the given instruction, if its pc flag is set, or else refers to the data flow of the closest previous instruction that updates the same state variable. For example, \texttt{36606800 ite 2 16606800 200 36606002} either selects the value of register \texttt{zero} (nid \texttt{200}) for updating the value of register \texttt{t0}, if \texttt{R0:addi t0,zero,0} is currently executing (nid \texttt{16606800}), or else refers to the \texttt{ite} expression for \texttt{SEGFL:ld t0,0(t0)} at \texttt{36606002} which may also update \texttt{t0}, and so on. Finally, the \texttt{next} value of registers such as \texttt{t0} at nid \texttt{205} is determined by lines whose nids begin with digit \texttt{6} such as \texttt{60000005 next 2 205 36606800 t0} which refers to the head of the \texttt{ite} cascade for \texttt{t0} at~\texttt{36606800}. Actual computation can be seen in \texttt{36605600 add 2 205 206} which adds the values of registers \texttt{t0} and \texttt{t1} (nid \texttt{206}) as instructed by \texttt{A1:add t0,t0,t1}. Line \texttt{36604003 read 2 20000000 36604001} models the memory \texttt{read} at address \texttt{gp-16} as instructed by \texttt{A0:ld t0,-16(gp)}.

Fig.~\ref{fig:control} shows a conditional control-flow fragment of the BTOR2 model generated from the running example in Fig.~\ref{fig:source}. For example, line \texttt{56606802 next 1 16606800 56606801} sets the pc flag of \texttt{R0:addi t0,zero,0} (nid \texttt{16606800}) either if \texttt{beq t0,zero,8[R0]} is currently executing (its pc flag at nid \texttt{16603600} is set) and the value of register \texttt{t0} is equal to 0 (line at nid \texttt{36603600}), or else if \texttt{sd t0,-8(s0)} is currently executing (pc flag at nid \texttt{16606400} and the line at nid \texttt{56606801}). If \texttt{beq t0,zero,8[R0]} is currently executing and the value of register \texttt{t0} is not equal to 0 (line at nid \texttt{36603601}) then the pc flag of \texttt{A0:ld t0,-16(gp)} is set (line at \texttt{56604001}). Note that only the translation of \texttt{beq} as well as \texttt{jal} and \texttt{jalr} instructions results in BTOR2 code that \emph{connects} control and data flow. While there are only finitely many jump targets with \texttt{jal} and \texttt{jalr} instructions (RISC-U binaries are static), \texttt{beq} instructions remain as the only source of path explosion with \texttt{read} system calls being the only source of data explosion, as shown next.

Lastly, Fig.~\ref{fig:system} shows a system call and \texttt{bad} state fragment of the BTOR2 model generated for the running example. In particular, it shows how input flows into the model through a \texttt{read} system call, how a potential segmentation fault is detected as \texttt{bad} state, and how main memory is written to, in this case, through the \texttt{read} system call. One-byte input (nid \texttt{81}) is unsigned-extended to a 64-bit memory word (nid \texttt{91}) and then flows via an \texttt{ite} cascade (head at nid \texttt{42000020}) to a \texttt{write} operator (nid \texttt{42000022}). The address for the \texttt{write} operator is \texttt{a1 + a0} (nid \texttt{42000021}) where register \texttt{a0} (nid \texttt{210}) is a cursor over the write buffer that was originally passed to the \texttt{read} system call in register \texttt{a1} (nid \texttt{211}). A potential segmentation fault, such as through \texttt{SEGFL:ld t0,0(t0)}, is detected in line \texttt{80000009 bad 80000008 b2} if there is any memory access at an address above the heap (bump pointer value of the \texttt{brk} system call at nid \texttt{45000001}) and below the stack (pointer nid \texttt{202}) where nid \texttt{44000001} is the head of an \texttt{ite} cascade over all addresses used in \texttt{read} and \texttt{write} operators. There are similar \texttt{bad} states for other unsafe parts of memory. All of main memory is updated by a single \texttt{next} (nid \texttt{70000000}) that refers to the head of an \texttt{ite} cascade at \texttt{42000028} over all \texttt{write} operators in the model.

There is a fun fact that we like to mention: running BEATOR is fast and since it is written in C* it can actually model itself and the rest of the unicorn toolchain written in C* as well. The BTOR2 model of the whole C* toolchain takes less than a second to build and is around 4MB.

\subsection{BEATOR Loves QUBOT Loves BEATOR}

Dead code elimination, constant propagation, and \emph{bounded memory modeling} are all effective translation techniques in reducing the number of qubits in a QUBO model. Moreover, they can be done by BEATOR or by QUBOT. We have therefore added support of all three to BEATOR. By bounded memory modeling we mean modeling a segmenting MMU and RAM in a bounded number of 64-bit bitvectors, one for each memory word in RAM, resulting in BTOR2 models that do not contain any arrays of bitvectors and thus any \texttt{read} and \texttt{write} operators anymore. The segmenting MMU bounds the size of the data, heap, and stack segments and then maps them, using the \texttt{slice} operator of BTOR2, from the 64-bit RISC-U virtual address space to a minimal physical address space. For the running example, a 4-bit physical address space is sufficient with only 12 memory words actually being accessed: 2 data words, 1 heap word, and 9 stack words while only the heap word (\texttt{*x}) and 1 stack word (\texttt{a}) are ever updated, see again Fig.~\ref{fig:memory} for the memory layout.

The RAM option takes the MMU option further by mapping each access to a memory word onto the 64-bit RAM bitvectors. RAM removes the need for arrays of bitvectors but increases BEATOR's complexity to $\Theta(m\cdot|P|)$ where $m$ is the size of memory accessed by $P$. However, with RAM modeling already done in BEATOR, QUBOT's complexity effectively remains the same, maintaining the toolchain's complexity. The MMU option can be used without the RAM option and only reduces memory size without changing BEATOR's complexity. We have also implemented a third, combined MMURAM option in BEATOR that mimics QUBOT's implementation which maps virtual addresses directly to the 64-bit RAM bitvectors.

Constant propagation in BEATOR runs in $\mathcal{O}(n)$ time by executing the first $n$ RISC-U instructions of a binary until reaching the first \texttt{read} call. Then, BEATOR eliminates dead code and snapshots the machine state including the heap as initial state in the BTOR2 model. QUBOT takes advantage of the additional $\mathcal{O}(n)$ time, maintaining the toolchain's complexity, and propagates constants beyond the first \texttt{read} call. However, propagating constants until the first \texttt{read} call is faster in BEATOR. Quantitative details are in Section~\ref{sec:exp}.

\section{Quantum Modeling with QUBOT}
\label{sec:quantum}

We introduce, by continuing to use the running example, the BTOR2-to-QUBO translator called \emph{QUBOT} written in Python in around 3-KLOC. A QUBO model is a binary quadratic (BQ) function, which QUBOT generates from a BTOR2 model. A BQ function is a quadratic function from binary variables to positive real values and 0 (ground energy on a quantum annealer). Constant factors are signed real values that are eventually normalized before solving the function. For example, the following BQ function encodes the logic gate for $NOT(x)=y$:
$$
NOT_{BQ}(x,y)=2-2x-2y+4xy
$$
Try evaluating the function to see when it reaches ground energy! Instead of generating a file, QUBOT represents a QUBO model as upper-diagonal matrix of constant factors in memory using a Python library by D-Wave Systems\textsuperscript{\ref{fot:dwave}}. The diagonal of the matrix represents linear factors such as $-2x$ and the upper part represents bi-linear factors such as $4xy$. Quadratic factors of binary variables such as $-2x^2$ are actually linear factors\footnote{thus BQ functions are actually binary bi-linear functions} because $-2x^2=-2x$. The model may be output in various formats including visualizations, see Fig.~\ref{fig:qubo}. QUBOT runs in $\mathcal{O}(n^2)$ time and space such that:

\begin{proposition}
\label{prop:qubot}
Let $B$ be the BTOR2 model generated by the C* compiler and BEATOR for a given C* program and machine state $M$. Moreover, let $n$ be a bound on number of state transitions. Then, for all model input $I$, the \texttt{bad} state in $B$ representing $M$ is reachable on $I$ in no more than $n$ state transitions if and only if the QUBO model generated by QUBOT for $B$ and $n$ evaluates to 0 on $I$.
\end{proposition}

Thus, with more qubits on a quantum annealer than variables in the QUBO model, quantum annealing the model may reach 0 (ground) energy for all model input $I$ that is part of the ground state if and only if the \texttt{bad} state is reachable in $B$ on $I$ in no more than $n$ state transitions. If there are not enough qubits on the machine, annealing may still work using a hybrid solver~\cite{Kurowski20}. The size of the model comes down to $\mathcal{O}(n)$ if memory consumption of $P$ is bounded by a constant.

\begin{table}
\centering
\begin{tabular}{|l|c|c|}
\hline
BEATOR2 ($\subset$ BTOR2) & \multicolumn{2}{c|}{\#binary variables or \#qubits} \\
\hline
                          & QUBOT & QUARC \\
\hline\hline
  \texttt{sort bitvec}, \texttt{array} & \multicolumn{2}{c|}{0} \\
\hline\hline
\multicolumn{3}{|l|}{Combinational Operators (related to SMT-LIB~\cite{SMTLIB})} \\
\hline\hline
  \texttt{constd} & \multicolumn{2}{c|}{0} \\
\hline
  \texttt{add}, \texttt{sub} & \multicolumn{2}{c|}{$\mathcal{O}(w)$} \\
  \hline
  \texttt{mul} & $\mathcal{O}(w^2)$ & $\mathcal{O}(w)$\\
  \texttt{udiv}, \texttt{urem} & $\mathcal{O}(w^2)$ constrained circuit & - \\
  \texttt{uext} & 0 &  $\mathcal{O}(w)$ \\
  \hline
  \texttt{slice} & \multicolumn{2}{c|}{0} \\
\hline
  \texttt{ite}, \texttt{and}, \texttt{not}, \texttt{eq} & \multicolumn{2}{c|}{$\mathcal{O}(w)$} \\
\hline
  \texttt{neq}, \texttt{ult(e)}, \texttt{ugt(e)} & \multicolumn{2}{c|}{$\mathcal{O}(w)$} \\
\hline
  \texttt{read}, \texttt{write} & $\mathcal{O}(m\cdot w)$ & - \\
\hline\hline
\multicolumn{3}{|l|}{Sequential Operators (not in SMT-LIB~\cite{SMTLIB})} \\
\hline\hline
  \texttt{state}, \texttt{input} & \multicolumn{2}{c|}{$\mathcal{O}(w)$ variables} \\
  \texttt{init} &  \multicolumn{2}{c|}{0} \\
  \hline
  \texttt{next} & $\mathcal{O}(m\cdot w^2\cdot|P|)$ &  $\mathcal{O}(m\cdot w\cdot|P|)$ \\
\hline
  \texttt{bad} & \multicolumn{2}{c|}{$\mathcal{O}(w\cdot|P|)$ circuit} \\
\hline
\end{tabular}
\caption{Size of QUBO model and quantum circuit components (in number of binary variables or qubits) with respect to BEATOR2 declarations and operators where $|P|$ is original C* program size, if applicable, $w$ is machine word size, and $m$ is memory size ($|P|$ and $w$ are runtime constants)}
\label{tbl:qubo}
\end{table}

The key challenge in the translation is to encode reachability in a state machine (BTOR2 model) in a stateless BQ function (QUBO model). Before learning how QUBOT translates our running example, let us focus on the fact that any combinational circuit can be represented by a BQ function:

\begin{proposition}
Let $x$, $y$, and $z$ be binary variables and let $f$ be a function (logic gate) that maps $x$ and $y$ to $z$. There exists at least one BQ function $g$ over $x$, $y$, and $z$ such that $g(x,y,z)=0$ if and only if $f(x,y)=z$ for all $x$, $y$, and $z$ in $\{0,1\}$.
\end{proposition}

For example, for all $x$, $y$, and $z$ in $\{0,1\}$, $AND(x,y)=z$ if and only if $6z+2xy-4xz-4yz=0$. Even though $AND$ and $NOT$ are universal, we also use dedicated BQ functions for $NAND$, $OR$, and $AND(NOT(x),y)$ (inhibition) for encoding the combinational operators in BTOR2. Similar to SMT solvers such as Z3~\cite{Z3} and boolector~\cite{Boolector}, QUBOT uses (polynomial) bit blasting except for \texttt{udiv} and \texttt{urem} which are reduced to constraints over multiplication. However, since solving those constraints is expensive, QUBOT cannot efficiently validate QUBO models generated from \texttt{udiv} and \texttt{urem}. Bit blasting both operators is future work. Memory access through \texttt{read} and \texttt{write} is bit-blasted, see below for more details. Note that composing a function from multiple BQ functions that generally share variables is straightforward using addition since the composed function is again a BQ function from binary variables to positive values and 0. Table~\ref{tbl:qubo} provides a summary of the size of the BQ functions for all BTOR2 operators in terms of the size $w$ of a machine word (here 64) and the size $m$ of memory. The impact of the size $|P|$ of the original C* program $P$ is also mentioned, see below for more on that.

The key idea of QUBOT is to model, given a bound $n$, the input $I$ and state variables $S$ in a BTOR2 model in $\mathcal{O}(n)$ qubits (binary variables) that are biased and entangled with another set of $\mathcal{O}(n)$ qubits that represent the output of the combinational circuits for control and data flow on $I$ and $S$. At least logically that model is then duplicated $n$ times where, for all $0\leq i<n$, duplicate $D_i$ is entangled with duplicate $D_{i+1}$ such that each qubit $o$ in $D_i$ that represents an output of a combinational circuit in the BTOR2 model is used in~$D_{i+1}$ as qubit that represents the part of the state variable in the BTOR2 model which $o$ updates~\cite{ASPLOS19}. Input variables are introduced in each duplicate as uninitialized state variables. After that, constants and initial values of initialized state variables are propagated at bit level (not word level) through the model.

This is the logic. In reality, QUBOT builds the model incrementally propagating constants as soon as possible to keep the model from growing unnecessarily. Instead of duplicating $D_i$, the model for $D_{i+1}$ is built from scratch, one combinational circuit per \texttt{next} operator at a time, immediately followed by propagating constants through that circuit. However, there is an issue: a qubit $o$ that represents an output of a circuit must not be replaced by a constant before all its future entanglements (uses) are known and have been modeled, which in the worst case may happen at $i=n-1$. The reason is that $o$ might represent part of a state variable not just in $D_{i+2}$ but also later until that part of the state variable is updated again by a qubit other than $o$. In short, state is memory (registers), and memory that is not updated keeps its value. Thus QUBOT remembers each such qubit $o$ and its possibly known constant value until it can be applied safely.

Incremental constant propagation in QUBOT is highly effective---all state variables including all pc flags are initialized---but it is certainly not exhaustive. Static analysis techniques other than constant propagation for further reducing the number of qubits may apply as well but remain future work. One such opportunity in the running example is the subtraction in \texttt{a = a - 1} in the body of the \texttt{while} loop. If the loop condition \texttt{a > '0'} was known at the time of subtraction, the bit resolution of the data flow through variable \texttt{a} could remain at 8 bits rather than 64 bits because \texttt{a - 1} never overflows into the 56 more significant bits. However, through mere constant propagation QUBOT is currently unable to perform that reasoning. Nevertheless, note that constant propagation enables QUBOT to validate models on known inputs efficiently, see Section~\ref{sec:exp}.

Let us now go through the running example from QUBOT's perspective. Given a BTOR2 model such as the model for our running example, QUBOT only looks for \texttt{next} and \texttt{bad} operators. For each line with a \texttt{next} operator, such as \texttt{60000005 next 2 205 36606800 t0} for data flow into registers in Fig.~\ref{fig:data} and \texttt{70000000 next 3 20000000 42000028} for data flow into memory in Fig.~\ref{fig:system}, and \texttt{56606802 next 1 16606800 56606801} for control flow in Fig.~\ref{fig:control}, QUBOT builds the BQ functions for the referenced (here, 64-bit and 1-bit bitvector) combinational circuits recursively, here nids \texttt{36606800}, \texttt{42000028}, and \texttt{56606801}, respectively. Lines with \texttt{state}, \texttt{input}, and \texttt{constd} operators terminate the recursion. When building $D_0$, QUBOT also looks for lines with \texttt{init} operators that initialize the encountered \texttt{state} variables and then, together with the encountered \texttt{constd} constants, performs constant propagation as described above. For each line with a \texttt{bad} operator, such as \texttt{80000009 bad 80000008 b2} for detecting potential segmentation faults in Fig.~\ref{fig:system}, QUBOT again builds the BQ function for the referenced (1-bitvector) combinational circuit recursively, here nid \texttt{80000008}, and then constrains the BQ function in an $OR$ circuit over all BQ functions generated for lines with \texttt{bad} operators to 1 (true).

Updating memory through the line \texttt{70000000 next 3 20000000 42000028} in Fig.~\ref{fig:system} works just like any other line with a \texttt{next} operator, except that the referenced (array) combinational circuit involves \texttt{read} and \texttt{write} operators. The issue with these operators is that QUBOT generates a BQ function that compares the referenced (64-bit bitvector) address, such as nid \texttt{42000021} in the line \texttt{42000022 write 3 20000000 42000021 42000020}, with all addresses of the entire memory to identify the correct memory word, which is then read or written, here with the output of the referenced (64-bit bitvector) combinational circuit (nid \texttt{42000020}).

In order to address scalability, QUBOT maps virtual addresses to a minimal physical address space at translation time, similar to the combined MMURAM option in BEATOR. Here, BEATOR  generating \texttt{bad} states for detecting segmentation faults in the BTOR2 model frees QUBOT from checking again. For the running example, a 4-bit physical address space for 12 memory words is sufficient, see again Fig.~\ref{fig:memory} for the memory layout. Most importantly, constant propagation in QUBOT reduces BQ functions modeling memory access via constant addresses to BQ functions modeling register access, effectively making their size independent of memory size. While we discussed QUBOT's perspective on a BTOR2 model generated by the unicorn toolchain, we should mention that QUBOT is able to handle any model within the BEATOR2 subset of BTOR2 and thus enables bounded model checking in general within the theory of bitvectors and arrays of bitvectors on a quantum annealer.

In sum, given a C* program $P$ and a bound $n$, the unicorn toolchain takes $\mathcal{O}(n^2)$ time to generate a QUBO model for $P$ where generating each $D_i$ for $0\leq i<n$ takes $\mathcal{O}(n)$ time. Translation time reduces to $\mathcal{O}(n)$ if memory consumption of $P$ is bounded by a constant. Note that the size $|P|$ of the C* program $P$ as well as the size $w$ of a machine word are additional linear and quadratic factors, respectively, in QUBOT's time and space complexity which we nevertheless ignore here since $|P|$ and $w$ are both runtime constants.

\section{Quantum Modeling with QUARC}
\label{sec:quarc}

We introduce the BTOR2-to-OpenQASM translator called \emph{QUARC} written in Python in around 1.5-KLOC. OpenQASM~\cite{qasm} is an open standard for specifying quantum circuits with tool support for validation, simulation, and deployment on real gate-model quantum computers. QUARC generates quantum circuits using X (logic not), CNOT (controlled not) gates\footnote{The inputs of a CNOT gate are a control and a target qubit. If the control qubit is 1, then the target qubit is flipped.}, and a generalization of CNOT gates that can have an arbitrary number of control qubits known as Toffoli gates. QUARC processes BTOR2 models with the same time and space complexity as QUBOT applying the same recursion that begins at \texttt{next} and \texttt{bad} operators and terminates at \texttt{state}, \texttt{input}, and \texttt{constd} operators:

\begin{proposition}
\label{prop:quarc}
Let $B$ be the BTOR2 model generated by the C* compiler and BEATOR for a given C* program and machine state $M$. Moreover, let $n$ be a bound on number of state transitions. Then, for all model input $I$, the \texttt{bad} state in $B$ representing $M$ is reachable on $I$ in no more than $n$ state transitions if and only if the quantum circuit generated by QUARC for $B$ and $n$ outputs 1 on $I$.
\end{proposition}

Quantum circuits generated by QUARC can be used as oracles with Grover's algorithm, a quantum search algorithm that provides quadratic speedup over brute force search in terms of the number of elements the search space has~\cite{Grover96}. In our case, the search space is given by the size of a circuit's input space since non-input qubits have either a classical value or are transitively entangled with input qubits:

\begin{proposition}
\label{prop:grover_quarc}
Let $C$ be a quantum circuit generated by QUARC for a given C* program, a bound $n$ on number of state transitions, and a machine state $M$. If the program reads up to $i$ bits of input executing no more than $n$ machine instructions, then finding such input where $C$ outputs 1, that is, the program reaches $M$ takes $\mathcal{O}(\sqrt{2^i}n^2)$ time using $C$ as oracle of size $\mathcal{O}(n^2)$ with Grover's algorithm~\cite{Grover96}.
\end{proposition}

Even though QUBO models generated by QUBOT work in principle as oracles with Grover's algorithm as well, generating dedicated quantum circuits is more efficient in the number of qubits since quantum operators can be used to avoid full bit-blasting and reduce ancillae (qubits for storing intermediate calculations). For example, QUARC uses fewer qubits by a linear factor on word size than QUBOT for implementing multiplication by reusing qubits in reversible parts of the circuit, see Table~\ref{tbl:qubo}. Support of division and remainder is future work. Word extension can also be done without introducing new qubits through constant propagation but remains future work as well. For now, QUARC applies constant propagation to prune branches of \texttt{ite} operators, and to reduce the number of qubits involved in multiqubit gates since they are harder to implement and take longer to execute on real hardware. The main principled contribution of this paper is:

\begin{proposition}
\label{prop:quadratic}
The algorithmic time (space) complexity of an algorithm in classical computing is a linear (quadratic) upper bound on quantum space in number of qubits for running as well as symbolically executing the algorithm on quantum annealers and gate-model quantum computers.
\end{proposition}

\section{Experiments}
\label{sec:exp}

\begin{figure}
\centering
\includegraphics[width=0.9\columnwidth,keepaspectratio]{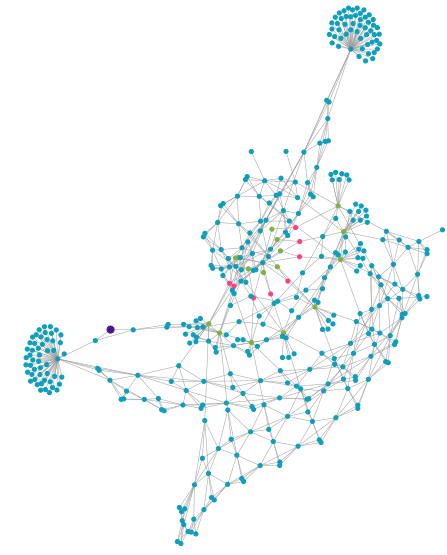}
\caption{QUBO model of the running example with 15 state transitions (input qubits are in pink, registers in green and pink, the big purple dot is a bad state).}
\label{fig:qubo}
\end{figure}

First of all, we managed to run real 64-bit and 32-bit C code on D-Wave's Advantage 5k-qubit quantum annealer successfully! This means that the quantum annealer was able to determine input that actually leads to bad states. In fact, the answers returned by the quantum annealer that achieved ground energy had the same value for \emph{all} binary variables as the ones our debugging tool produce for the specific inputs returned by the quantum annealer. While we had to replace the code below the assignment \texttt{a = *x;} in the running example with just one assignment \texttt{a = *(x + a);}, which may trigger a segmentation fault for all inputs other than \texttt{0}, we still feel that this is a breakthrough and, to the best of our knowledge, the first time C code, using dynamic memory allocation (as simple as it might be) and not targeting quantum annealers~\cite{HPEC19}, ran on a quantum computer through a fully automated toolchain.

\begin{table}[]
  \begin{tabular}{|c|c|c|c|c|}
    \hline
    \multirow{2}{*}{program} & \multirow{2}{*}{\begin{tabular}[c]{@{}c@{}}\#state \\ transitions\end{tabular}} & \multirow{2}{*}{bad states} & \multicolumn{2}{c|}{\#logic qubits (32-bit)} \\
  \cline{4-5}
   &  & & QUBOT & QUARC \\
  \hline \hline
  division-by-zero & 16 & division by zero & 5k & - \\ \hline
  memory-access-fail & 0 & \begin{tabular}[c]{@{}c@{}}memory access \\ in between data \\ and heap segments\end{tabular} & 0 & 0\\ \hline
  nested-if-else & 51 & \multirow{7}{*}{non-zero exit code} & 220k & 164k \\ \cline{1-2}
  nested-if-else-reverse & 50 & & 208k & 157k \\ \cline{1-2}
  return-from-loop & 35 & & 40k & 79k \\ \cline{1-2}
  simple-assignment & 31 & & 45k & 68k \\ \cline{1-2}
  simple-if-else & 45 & & 65k & 100k \\ \cline{1-2}
  simple-if-else-reverse & 43 & & 72k & 100k \\ \cline{1-2}
  simple-if-without-else & 44 & & 107k & 111k \\ \hline
  \end{tabular}
  \caption{C* programs in 32-bit version for validating unicorn by checking that QUBO models generated by QUBOT and quantum circuits generated by QUARC identify the same bad states, inputs, and state transitions as \texttt{btormc}~\cite{BTOR2}.}
  \label{tbl:test_programs}
\end{table}

Unicorn includes tools for debugging and validation of both, QUBO models generated by QUBOT and quantum circuits generated by QUARC. Debugging and validation works by propagating known input values through models and circuits to validate the output. Moreover, the values of qubits that represent bad states reveal which bad states actually occur. We used 32-bit versions of the C* programs listed in Table~\ref{tbl:test_programs} to debug and validate QUBO models and quantum circuits. For this purpose, BEATOR was configured to perform constant propagation until first input, down-scale the linear address space, and encode MMU and RAM access, to support both QUBOT and QUARC. Additionally, QUBO models in 32- and even 64-bit versions listed in Table~\ref{tbl:exp} were validated with all BEATOR configurations. Note that here validation is feasible because all programs have only a 1-byte input and thus only require checking 256 values each.

\begin{table}[h]
  \begin{tabular}{|c|c|c|c|}
    \hline
    QPU & release year & topology & \#qubits \\ \hline \hline
    DW2000 & 2017 & Chimera (C16) & 2048 \\ \hline
    Advantage & 2020 & Pegasus (P16) & 5640 \\ \hline
  \end{tabular}
  \caption{D-Wave's QPUs used in our experiments, the year they were released, the topology of the graph describing how qubits can be entangled, and the total number of qubits available.}
  \label{tbl:qpus}
\end{table}

\begin{table}[h]
  \centering
  \begin{tabular}{|c|c|c|c|c|}
    \hline
  wordsize & QPU & \#binary variables & \#physical qubits & time (s) \\ \hline \hline
  \multirow{2}{*}{32} & DW2000 & \multirow{2}{*}{348} & 1106 & 8.29 \\
                      & Advantage &                   & 589  & 3.12 \\ \hline
  \multirow{2}{*}{64} & DW2000 & \multirow{2}{*}{398} & 1126 & 12.16 \\
                      & Advantage &                   & 680 & 5.29  \\ \hline
  \end{tabular}
  \caption{The total number of variables needed and the time spent in seconds to find a minor embedding for the 2 latest D-Wave QPUs, given QUBO models with 348 and 398 variables.}
  \label{tbl:embedding}
\end{table}

Table~\ref{tbl:test_programs} also mentions the number of qubits that we would have needed in order to execute each of the programs symbolically. Here, QUBOT is effective in reducing the number of qubits for programs with very few state transitions because of its eager constant propagation. However, QUARC does a better job when constants cannot be further propagated since it needs, in terms of word size, even asymptotically fewer ancillae to represent each operator.


\begin{table}[]
  \begin{tabular}{|c|c|c|c|c|c|}
    \hline
  wordsize            & QPU                        & \#samples               & chain strength & \#minimum energy samples & minimum energy \\ \hline \hline
  \multirow{4}{*}{32} & \multirow{6}{*}{P16} & \multirow{3}{*}{7k} & 1.5 & 12 & 0 \\ \cline{4-6}
                      &  &  & 2 & \multirow{2}{*}{0}  & 2 \\ \cline{4-4} \cline{6-6}
                      &  &  & 1 & & 6                     \\ \cline{3-6}
                      &  & 10k & \multirow{3}{*}{1.5} & 9 & \multirow{3}{*}{0} \\ \cline{1-1} \cline{3-3}\cline{5-5}
  \multirow{2}{*}{64} &  & 7k &  & \multirow{2}{*}{1}   & \\ \cline{3-3}
                      &  & 10k &   &  & \\ \hline
  \multirow{4}{*}{32} & \multirow{5}{*}{C16}   & \multirow{3}{*}{7k} & 1.5 & \multirow{5}{*}{0} & \multirow{2}{*}{2} \\ \cline{4-4}
  & &  & 2 & & \\ \cline{4-4} \cline{6-6}
  & &  & 2.5  &  & 4  \\ \cline{3-4} \cline{4-4} \cline{6-6}
 & & \multirow{2}{*}{10k} & \multirow{2}{*}{1.5}  &  & \multirow{2}{*}{2} \\\cline{1-1}
  64  &  &  & & & \\ \hline
  \end{tabular}
    \caption{Shows how varying the number of samples and chain strength affects the number of low-energy solutions found. For example, the QUBO model for 32-bit machine words achives the highest number of solutions (12) that have 0 energy when 7k samples are taken on the QPU and the chain strength is set to 1.5.}
    \label{tbl:qa_results}
  \end{table}

\begin{figure}[H]
  \centering
  \includegraphics[width=0.6\columnwidth,keepaspectratio]{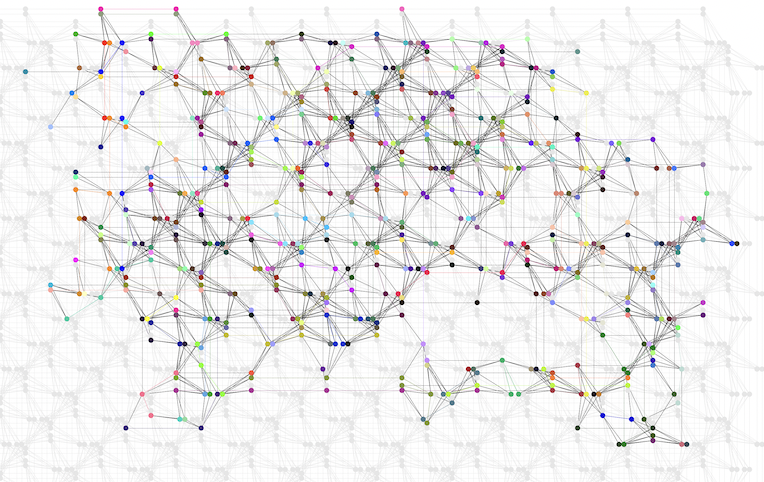}
  \caption{Minor embedding of the modified running example in a QPU with a Pegasus topology. Nodes with the same color represent the same binary variable.}
  \label{fig:embedding}
  \end{figure}

  \begin{figure}[H]
    \centering
    \includegraphics[width=0.7\columnwidth,keepaspectratio]{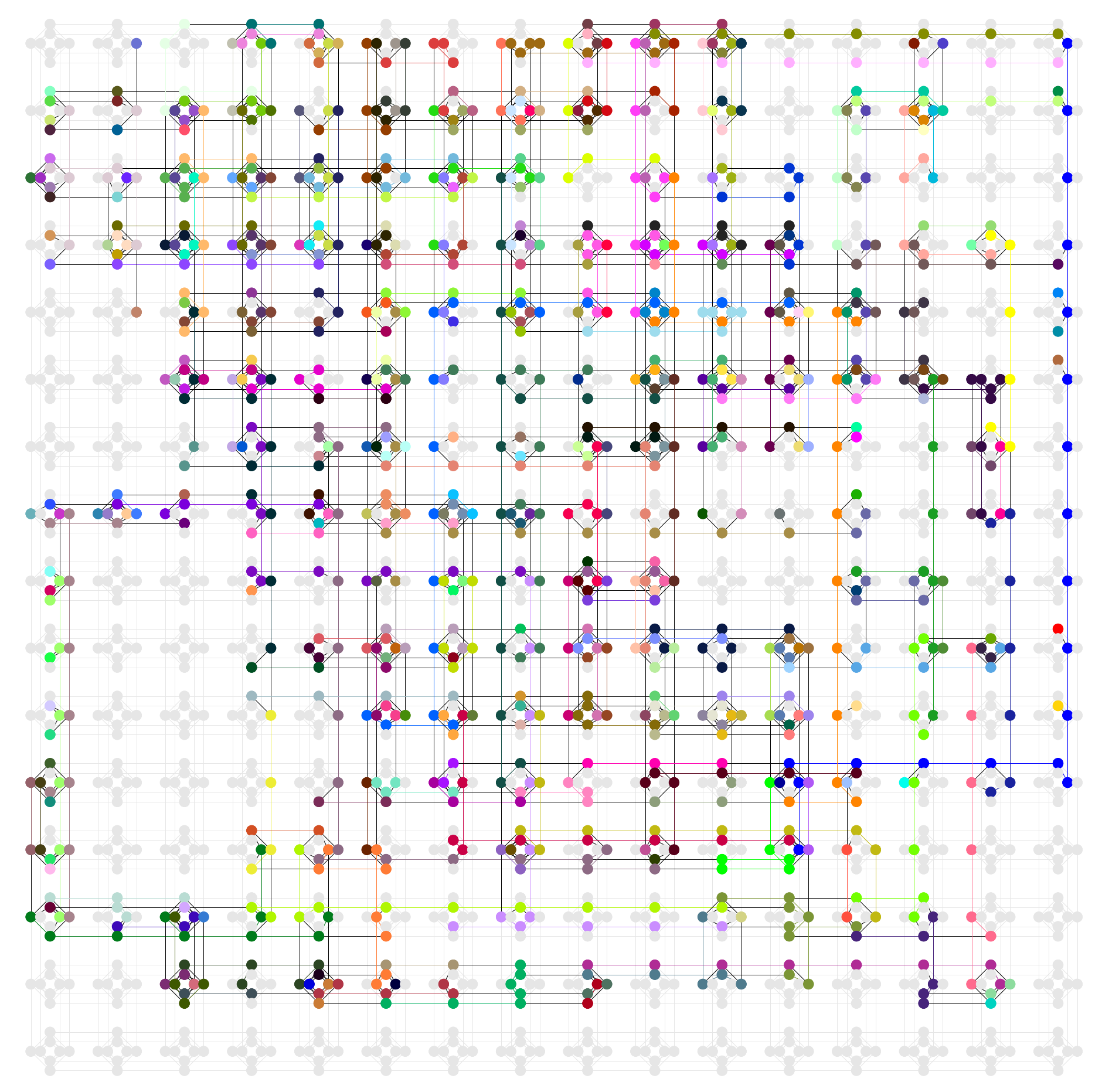}
    \caption{Minor embedding of the modified running example in a QPU with a Chimera topology. Nodes with the same color represent the same binary variable.}
    \label{fig:chimera_embedding}
  \end{figure}

We executed the running example varying only basic settings on the two latest QPUs provided by D-Wave (See Table~\ref{tbl:qpus}). Tables~\ref{tbl:embedding} and~\ref{tbl:qa_results} reveal that DW2000Q\_6 fails to find solutions, doubles ancillae, and more than doubles the time to find a minor embedding\footnote{We used the default algorithm by D-Wave~\cite{Cai14}.} than the newer Advantage QPU. We vary the number of samples and the strength by which physical variables (chain strength) are connected to represent a logic variable. On Advantage, we got a maximum of 12 answers after sampling 7k times the 32-bit version and 1 answer for the 64-bit version. In practice, the annealing time is just 20$\mu$s per sample but additional time is needed among samples, and to program and read the QPU. Still, experiments show that in the time span between these two QPUs (3 years), there is significant improvement. Also, the more qubits and connectivity a QPU has, the easier it is to find minor embeddings. Fig.~\ref{fig:embedding} shows that only a small portion of the QPU is utilized in Advantage when compared to the embedding done in DW\_2000 shown in Fig.~\ref{fig:chimera_embedding} for the same QUBO model.

Fig.~\ref{fig:qubo} visualizes the QUBO model with 398 variables of the 64-bit version of the example executed on the quantum annealer that needs 15 state transitions to raise an exception. Vertices are the variables of the QUBO model. An edge means that there is a non-zero bi-linear factor between two variables. All state transitions may generate bad states. However, constant propagation determines a constant value for all except for one (big purple dot). This figure suggest there is a lot of potential improvement that can be done in terms of number of variables needed since only 8 variables are input (pink vertices), 24 of them belong to registers, 8 are in pink because they are inputs too while the rest (16 vertices) are in green. Input variables are reused in registers, and in main memory.

\begin{figure}
  \centering
  \includegraphics[width=0.8\columnwidth,keepaspectratio]{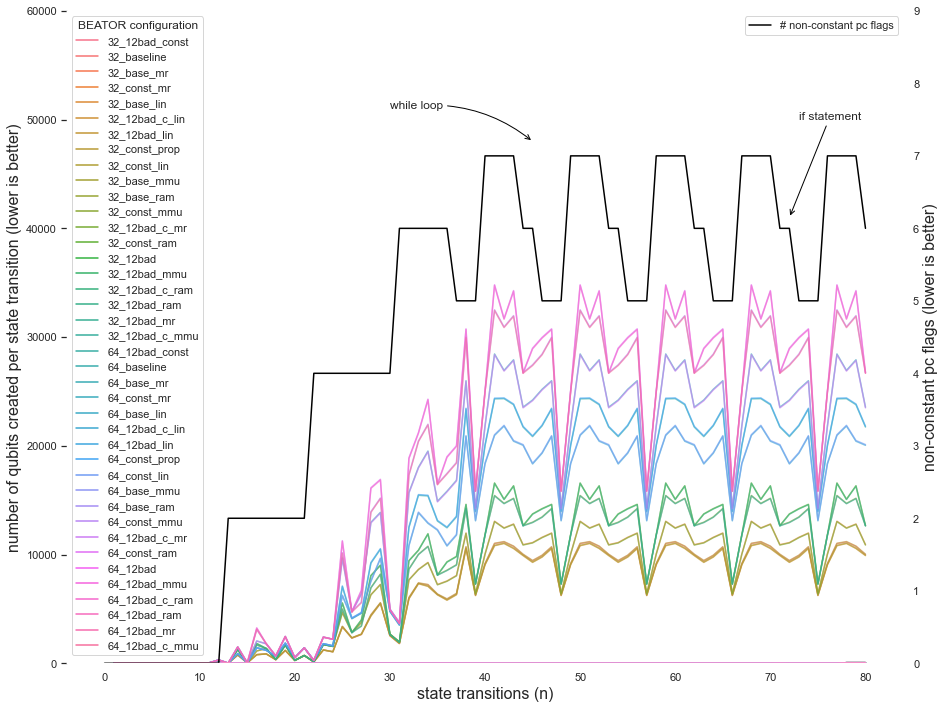}
  \caption{Left y-axis: number of qubits per state transition (x-axis) for the running example (no growth is linear growth in total qubit count, data on 32-bit models is prefixed 32, the outliers at the top are BEATOR's RAM configurations); right y-axis: number of non-constant pc flags (number of RISC-U instructions executed simultaneously, showing (linear) impact of the \texttt{while} loop and \texttt{if} statement on qubit count)}
  \label{fig:exp}
  \end{figure}

Fig.~\ref{fig:exp} shows the number of qubits created by QUBOT per state transition of the unmodified running example (Fig.~\ref{fig:source}) since it contains a more interesting control and data flow that can show interesting effects of constant propagation. The number of RISC-U instructions executed simultaneously (non-constant pc flags) shows the (linear) impact of control flow on qubit count, with the \texttt{while} loop and \texttt{if} statement showing up. Generally, the more effective BEATOR configurations result in slower qubit growth in QUBOT. Gaining quantum advantage essentially requires improving that data.

\begin{table}
\centering
\begin{tabular}{|l|c|c|c|}
\hline
BEATOR Config. & KB & Time (s) & \#Qubits (k) \\
\hline\hline
DEFAULT     & 51[50] & 7+3[4+1] & 180[92] \\
\hline
LINEAR      & 53[52] & 7+3[4+1] & 164[91] \\
\hline
MMU         & 78[77] & 9+4[5+2] & 204[106] \\
\hline
RAM         & 136[135] & 9+4[5+2] & 221[115] \\
\hline
MMURAM      & 100[99] & 7+3[4+2] & 197[100] \\
\hline
CONST-PROP  & 33[32] & \textbf{3+1[2+1]} & 140[70] \\
\hline
CONST-LIN   & 34[33] & 6+2\textbf{[2+1]} & 126[70] \\
\hline
CONST-MMU   & 47[46] & 5+2\textbf{[2+1]} & 188[98] \\
\hline
CONST-RAM   & 69[68] & 5+2\textbf{[2+1]} & 201[105] \\
\hline
CONST-MR    & 52[51] & 4+2[3+1] & 173[89] \\
\hline
12BAD       & 50[49] & 8+3[4+1] & 180[91] \\
\hline
12BAD-LIN   & 51[50] & 7+3[4+2] & 163[90] \\
\hline
12BAD-MMU   & 77[75] & 8+3[6+2] & 204[106] \\
\hline
12BAD-RAM   & 135[133] & 9+3[5+2] & 220[114] \\
\hline
12BAD-MR    & 99[98] & 8+3[4+2] & 196[99] \\
\hline
12BAD-CONST & \textbf{31[30]} & \textbf{3+1[2+1]} & 138[70] \\
\hline
12BAD-C-LIN & 32[31] & \textbf{3+1[2+1]} & \textbf{125[69]} \\
\hline
12BAD-C-MMU & 46[45] & 5+3\textbf{[2+1]} & 187[97] \\
\hline
12BAD-C-RAM & 68[67] & 4+2\textbf{[2+1]} & 201[104] \\
\hline
12BAD-C-MR  & 51[49] & 4+2\textbf{[2+1]} & 172[88] \\
\hline
\end{tabular}
\caption{BTOR2 model size in kilobyte (KB), QUBOT translation time in seconds (modeling time + constant-propagation time), number of generated qubits in thousands, 32-bit results in brackets (obtained with running example)}
\label{tbl:exp}
\end{table}

Improvements may come from both BEATOR and QUBOT. Table~\ref{tbl:exp} shows, for the running example, BTOR2 model size in kilobyte (KB), QUBOT translation time in seconds (model build time + constant propagation time), number of generated qubits in thousands, and 32-bit results in brackets. We interpret the data as follows: (1)~the lowest QUBOT translation time and qubit count is reached if BEATOR, in decreasing order of efficacy, propagates constants until the first input is read (\underline{\underline{C}ONST}-PROP), down-scales 64-bit virtual addresses to 29-bit [30-bit] linear addresses (LINEAR,LIN), and only generates bad states for segmentation faults (12BAD), (2)~modeling MMU and RAM is more effective if done by QUBOT (on bit level versus on word level in BEATOR (MMU,RAM,\underline{M}MU\underline{R}AM)), and (3)~32-bit models have, as expected, roughly half the qubit count of 64-bit models.

Constant propagation until first input is read is more effective in BEATOR than in QUBOT because the resulting program state is generally smaller. However, after that, constant propagation in QUBOT is highly effective, for example: 12BAD-C-LIN for 32-bit code results in around 3 million qubits without constant propagation in QUBOT. Downscaling to smaller linear address spaces is only slightly more effective in BEATOR than constant-propagating unused addressing bits in QUBOT while reducing the number of bad states in BEATOR has almost no effect. MMU and RAM modeling with constant propagation on bit level is generally more effective in QUBOT while support of 32-bit models can only be done in the toolchain before QUBOT.

\section{Conclusions and Future Work}
\label{sec:conc}

We have shown how to encode symbolic execution of RISC-V machine code compiled from C code in (1)~finite state machines over the theory of bitvectors and arrays of bitvectors, (2)~quadratic unconstrained binary optimization (QUBO) models, and (3)~quantum circuits, enabling not just bounded model checkers but also quantum annealers and gate-model machines to execute arbitrary code symbolically.

Our construction implies that the algorithmic complexity of any classical algorithm written in a Turing-complete programming language polynomially bounds the number of quantum bits that are required to run the algorithm on a quantum computer. In particular, any classical algorithm $A$ that runs in $\mathcal{O}(f(n))$ time and $\mathcal{O}(g(n))$ space requires no more than $\mathcal{O}(f(n)\cdot g(n))$ quantum bits to run on a quantum computer. With $\mathcal{O}(1)\leq\mathcal{O}(g(n))\leq\mathcal{O}(f(n))$ for all $n$, the quantum bits required to run $A$ may therefore not exceed $\mathcal{O}(f(n)^2)$ and may come down to $\mathcal{O}(f(n))$ if memory consumption is bounded by a constant.

There are numerous directions for principled future work where outperforming, that is, gaining quantum advantage over classical symbolic execution and bounded model checking is at the top of the challenges. So far, we have dealt with the fundamental tradeoff between translation and execution complexity by putting most hard, that is, exponential work into execution. Any static analysis technique applied at translation time that scales to the size of our models and circuits is likely to produce significant progress in reducing the amount of qubits required to solve problems in practice. QUBO models and quantum circuits play a key role in programming quantum annealers and gate-model machines but may require more attention to be established as key abstraction from quantum physics, similar to the role of Boolean algebra in classical computing.

There are also numerous technical challenges of which we list just a few. Besides the obvious such as full C, RISC-V, and BTOR2 support, there is also need for support of floating-point arithmetic, more efficient handling of symbolic memory addresses, detection of properties other than safety, and more advanced system call handling that covers incorrect use of system calls.

\section{Acknowledgements}
We thank Armin Biere for inspiring us to build BEATOR! This work has been supported by the Czech Ministry of Education, Youth and Sports from the Czech Operational Programme Research, Development, and Education, under grant agreement No.~CZ.02.1.01/0.0/0.0/15\_003/0000421, and the European Research Council (ERC) under the European Union’s Horizon 2020 research and innovation programme, under grant agreement No.~695412.

\bibliographystyle{ACM-Reference-Format}
\bibliography{paper}


\begin{thebibliography}{60}


\ifx \showCODEN    \undefined \def \showCODEN     #1{\unskip}     \fi
\ifx \showDOI      \undefined \def \showDOI       #1{#1}\fi
\ifx \showISBNx    \undefined \def \showISBNx     #1{\unskip}     \fi
\ifx \showISBNxiii \undefined \def \showISBNxiii  #1{\unskip}     \fi
\ifx \showISSN     \undefined \def \showISSN      #1{\unskip}     \fi
\ifx \showLCCN     \undefined \def \showLCCN      #1{\unskip}     \fi
\ifx \shownote     \undefined \def \shownote      #1{#1}          \fi
\ifx \showarticletitle \undefined \def \showarticletitle #1{#1}   \fi
\ifx \showURL      \undefined \def \showURL       {\relax}        \fi
\providecommand\bibfield[2]{#2}
\providecommand\bibinfo[2]{#2}
\providecommand\natexlab[1]{#1}
\providecommand\showeprint[2][]{arXiv:#2}

\bibitem[Abyaneh et~al\mbox{.}(2018)]%
        {MoreVMs18}
\bibfield{author}{\bibinfo{person}{A.S. Abyaneh}, \bibinfo{person}{S. Bauer},
  \bibinfo{person}{C.M. Kirsch}, \bibinfo{person}{P. Mayer},
  \bibinfo{person}{C. M{\"o}sl}, \bibinfo{person}{C. Poncelet},
  \bibinfo{person}{S. Seidl}, \bibinfo{person}{A. Sokolova}, {and}
  \bibinfo{person}{M. Widmoser}.} \bibinfo{year}{2018}\natexlab{}.
\newblock \showarticletitle{Selfie: Towards Minimal Symbolic Execution}. In
  \bibinfo{booktitle}{\emph{Online Proc. Workshop on Modern Language Runtimes,
  Ecosystems, and VMs (MoreVMs)}}.
\newblock


\bibitem[Abyaneh and Kirsch(2021)]%
        {ASE21}
\bibfield{author}{\bibinfo{person}{A.S. Abyaneh} {and} \bibinfo{person}{C.M.
  Kirsch}.} \bibinfo{year}{2021}\natexlab{}.
\newblock \showarticletitle{ASE: A Value Set Decision Procedure for Symbolic
  Execution}. In \bibinfo{booktitle}{\emph{Proc. IEEE/ACM International
  Conference on Automated Software Engineering (ASE)}}.
  \bibinfo{publisher}{IEEE/ACM}.
\newblock


\bibitem[Adachi and Henderson(2015)]%
        {Adachi15}
\bibfield{author}{\bibinfo{person}{Steven~H. Adachi} {and}
  \bibinfo{person}{Maxwell~P. Henderson}.} \bibinfo{year}{2015}\natexlab{}.
\newblock \bibinfo{title}{Application of Quantum Annealing to Training of Deep
  Neural Networks}.
\newblock
\newblock
\showeprint[arxiv]{1510.06356}~[quant-ph]


\bibitem[Barrett et~al\mbox{.}(2010)]%
        {SMTLIB}
\bibfield{author}{\bibinfo{person}{C. Barrett}, \bibinfo{person}{A. Stump},
  {and} \bibinfo{person}{C. Tinelli}.} \bibinfo{year}{2010}\natexlab{}.
\newblock \showarticletitle{The SMT-LIB Standard: Version 2.0}. In
  \bibinfo{booktitle}{\emph{Proc. International Workshop on Satisfiability
  Modulo Theories}}, \bibfield{editor}{\bibinfo{person}{A.~Gupta} {and}
  \bibinfo{person}{D.~Kroening}} (Eds.).
\newblock


\bibitem[Bass et~al\mbox{.}(2021)]%
        {Gideon20}
\bibfield{author}{\bibinfo{person}{G. Bass}, \bibinfo{person}{M. Henderson},
  \bibinfo{person}{J. Heath}, {and} \bibinfo{person}{J. Dulny}.}
  \bibinfo{year}{2021}\natexlab{}.
\newblock \showarticletitle{Optimizing the Optimizer: Decomposition Techniques
  for Quantum Annealing}.
\newblock \bibinfo{journal}{\emph{Quantum Mach. Intell.}}
  (\bibinfo{year}{2021}).
\newblock
\urldef\tempurl%
\url{https://doi.org/10.1007/s42484-021-00039-9}
\showURL{%
\tempurl}


\bibitem[Battiti and Protasi(1997)]%
        {Battiti97}
\bibfield{author}{\bibinfo{person}{R. Battiti} {and} \bibinfo{person}{M.
  Protasi}.} \bibinfo{year}{1997}\natexlab{}.
\newblock \showarticletitle{Reactive Search, a History-Sensitive Heuristic for
  MAX-SAT}.
\newblock \bibinfo{journal}{\emph{ACM J. Exp. Algorithmics}}
  \bibinfo{volume}{2} (\bibinfo{date}{1} \bibinfo{year}{1997}),
  \bibinfo{pages}{2–es}.
\newblock
\showISSN{1084-6654}
\urldef\tempurl%
\url{https://doi.org/10.1145/264216.264220}
\showDOI{\tempurl}


\bibitem[Bian et~al\mbox{.}(2017)]%
        {Bian17}
\bibfield{author}{\bibinfo{person}{Z. Bian}, \bibinfo{person}{F. Chudak},
  \bibinfo{person}{W. Macready}, \bibinfo{person}{A. Roy}, \bibinfo{person}{R.
  Sebastiani}, {and} \bibinfo{person}{S. Varotti}.}
  \bibinfo{year}{2017}\natexlab{}.
\newblock \showarticletitle{Solving SAT and MaxSAT with a Quantum Annealer:
  Foundations and a Preliminary Report}. In \bibinfo{booktitle}{\emph{Frontiers
  of Combining Systems}}, \bibfield{editor}{\bibinfo{person}{C.~Dixon} {and}
  \bibinfo{person}{M.~Finger}} (Eds.). \bibinfo{publisher}{Springer},
  \bibinfo{address}{Cham}, \bibinfo{pages}{153--171}.
\newblock
\showISBNx{978-3-319-66167-4}


\bibitem[Biere et~al\mbox{.}(1999)]%
        {BMC}
\bibfield{author}{\bibinfo{person}{A. Biere}, \bibinfo{person}{A. Cimatti},
  \bibinfo{person}{E.M. Clarke}, {and} \bibinfo{person}{Y. Zhu}.}
  \bibinfo{year}{1999}\natexlab{}.
\newblock \showarticletitle{Symbolic Model Checking without {BDDs}}. In
  \bibinfo{booktitle}{\emph{{TACAS}}} \emph{(\bibinfo{series}{LNCS},
  Vol.~\bibinfo{volume}{1579})}. \bibinfo{publisher}{Springer},
  \bibinfo{pages}{193--207}.
\newblock


\bibitem[Bonet et~al\mbox{.}(2007)]%
        {Bonet07}
\bibfield{author}{\bibinfo{person}{M.L. Bonet}, \bibinfo{person}{J. Levy},
  {and} \bibinfo{person}{F. Manyà}.} \bibinfo{year}{2007}\natexlab{}.
\newblock \showarticletitle{Resolution for Max-SAT}.
\newblock \bibinfo{journal}{\emph{Artificial Intelligence}}
  \bibinfo{volume}{171}, \bibinfo{number}{8} (\bibinfo{year}{2007}),
  \bibinfo{pages}{606--618}.
\newblock
\showISSN{0004-3702}
\urldef\tempurl%
\url{https://doi.org/10.1016/j.artint.2007.03.001}
\showDOI{\tempurl}


\bibitem[Boothby et~al\mbox{.}(2020)]%
        {Boothby20}
\bibfield{author}{\bibinfo{person}{K. Boothby}, \bibinfo{person}{A.~D. King},
  {and} \bibinfo{person}{A. Roy}.} \bibinfo{year}{2020}\natexlab{}.
\newblock \bibinfo{title}{Fast clique minor generation in Chimera qubit
  connectivity graphs}.
\newblock
\newblock
\showeprint[arxiv]{1507.04774}~[cs.DM]


\bibitem[Burckhardt et~al\mbox{.}(2007)]%
        {CheckFence}
\bibfield{author}{\bibinfo{person}{Sebastian Burckhardt},
  \bibinfo{person}{Rajeev Alur}, {and} \bibinfo{person}{Milo M.~K. Martin}.}
  \bibinfo{year}{2007}\natexlab{}.
\newblock \showarticletitle{CheckFence: Checking Consistency of Concurrent Data
  Types on Relaxed Memory Models}.
\newblock \bibinfo{journal}{\emph{SIGPLAN Not.}} \bibinfo{volume}{42},
  \bibinfo{number}{6} (\bibinfo{date}{jun} \bibinfo{year}{2007}),
  \bibinfo{pages}{12–21}.
\newblock
\showISSN{0362-1340}
\urldef\tempurl%
\url{https://doi.org/10.1145/1273442.1250737}
\showDOI{\tempurl}


\bibitem[Cadar et~al\mbox{.}(2008)]%
        {KLEE}
\bibfield{author}{\bibinfo{person}{C. Cadar}, \bibinfo{person}{D. Dunbar},
  {and} \bibinfo{person}{D. Engler}.} \bibinfo{year}{2008}\natexlab{}.
\newblock \showarticletitle{KLEE: Unassisted and Automatic Generation of
  High-Coverage Tests for Complex Systems Programs}. In
  \bibinfo{booktitle}{\emph{Proc. USENIX Conference on Operating Systems Design
  and Implementation (OSDI)}}. \bibinfo{publisher}{USENIX Association},
  \bibinfo{pages}{209–224}.
\newblock


\bibitem[Cai et~al\mbox{.}(2014)]%
        {Cai14}
\bibfield{author}{\bibinfo{person}{J. Cai}, \bibinfo{person}{B. Macready},
  {and} \bibinfo{person}{A. Roy}.} \bibinfo{year}{2014}\natexlab{}.
\newblock \bibinfo{title}{A practical heuristic for finding graph minors}.
\newblock
\newblock
\urldef\tempurl%
\url{https://doi.org/10.48550/ARXIV.1406.2741}
\showDOI{\tempurl}


\bibitem[Chipounov et~al\mbox{.}(2011)]%
        {S2E}
\bibfield{author}{\bibinfo{person}{V. Chipounov}, \bibinfo{person}{V.
  Kuznetsov}, {and} \bibinfo{person}{G. Candea}.}
  \bibinfo{year}{2011}\natexlab{}.
\newblock \showarticletitle{S2E: A Platform for in-Vivo Multi-Path Analysis of
  Software Systems}. In \bibinfo{booktitle}{\emph{Proc. International
  Conference on Architectural Support for Programming Languages and Operating
  Systems (ASPLOS)}}. \bibinfo{publisher}{ACM}, \bibinfo{pages}{265–278}.
\newblock


\bibitem[Cross et~al\mbox{.}(2017)]%
        {qasm}
\bibfield{author}{\bibinfo{person}{Andrew~W. Cross}, \bibinfo{person}{Lev~S.
  Bishop}, \bibinfo{person}{John~A. Smolin}, {and} \bibinfo{person}{Jay~M.
  Gambetta}.} \bibinfo{year}{2017}\natexlab{}.
\newblock \bibinfo{title}{Open Quantum Assembly Language}.
\newblock
\newblock
\urldef\tempurl%
\url{https://doi.org/10.48550/ARXIV.1707.03429}
\showDOI{\tempurl}


\bibitem[Crosson and Harrow(2016)]%
        {Crosson16}
\bibfield{author}{\bibinfo{person}{E. Crosson} {and} \bibinfo{person}{A.~W.
  Harrow}.} \bibinfo{year}{2016}\natexlab{}.
\newblock \showarticletitle{Simulated Quantum Annealing Can Be Exponentially
  Faster Than Classical Simulated Annealing}. In
  \bibinfo{booktitle}{\emph{Proc. Annual Symposium on Foundations of Computer
  Science (FOCS)}}. \bibinfo{publisher}{IEEE}, \bibinfo{pages}{714--723}.
\newblock
\urldef\tempurl%
\url{https://doi.org/10.1109/FOCS.2016.81}
\showDOI{\tempurl}


\bibitem[Crosson and Lidar(2021)]%
        {Crosson21}
\bibfield{author}{\bibinfo{person}{E.~J. Crosson} {and} \bibinfo{person}{D.~A.
  Lidar}.} \bibinfo{year}{2021}\natexlab{}.
\newblock \showarticletitle{Prospects for Quantum Enhancement with Diabatic
  Quantum Annealing}.
\newblock \bibinfo{journal}{\emph{Nature Reviews Physics}} \bibinfo{volume}{3},
  \bibinfo{number}{7} (\bibinfo{date}{5} \bibinfo{year}{2021}),
  \bibinfo{pages}{466–489}.
\newblock
\showISSN{2522-5820}
\urldef\tempurl%
\url{https://doi.org/10.1038/s42254-021-00313-6}
\showDOI{\tempurl}


\bibitem[Date et~al\mbox{.}(2019)]%
        {Date2019}
\bibfield{author}{\bibinfo{person}{P. Date}, \bibinfo{person}{R. Patton},
  \bibinfo{person}{C. Schuman}, {and} \bibinfo{person}{T. Potok}.}
  \bibinfo{year}{2019}\natexlab{}.
\newblock \showarticletitle{Efficiently Embedding QUBO Problems on Adiabatic
  Quantum Computers}.
\newblock \bibinfo{journal}{\emph{Quantum Information Processing}}
  \bibinfo{volume}{18} (\bibinfo{date}{03} \bibinfo{year}{2019}).
\newblock
\urldef\tempurl%
\url{https://doi.org/10.1007/s11128-019-2236-3}
\showDOI{\tempurl}


\bibitem[de~Moura and Bj{\o}rner(2008)]%
        {Z3}
\bibfield{author}{\bibinfo{person}{L. de Moura} {and} \bibinfo{person}{N.
  Bj{\o}rner}.} \bibinfo{year}{2008}\natexlab{}.
\newblock \showarticletitle{Z3: An Efficient SMT Solver}. In
  \bibinfo{booktitle}{\emph{Tools and Algorithms for the Construction and
  Analysis of Systems}}, \bibfield{editor}{\bibinfo{person}{C.~R. Ramakrishnan}
  {and} \bibinfo{person}{Jakob Rehof}} (Eds.). \bibinfo{publisher}{Springer},
  \bibinfo{pages}{337--340}.
\newblock


\bibitem[Douglass et~al\mbox{.}(2015)]%
        {King15}
\bibfield{author}{\bibinfo{person}{A. Douglass}, \bibinfo{person}{A.~D. King.},
  {and} \bibinfo{person}{J. Raymond}.} \bibinfo{year}{2015}\natexlab{}.
\newblock \showarticletitle{Constructing SAT Filters with a Quantum Annealer}.
  In \bibinfo{booktitle}{\emph{Theory and Applications of Satisfiability
  Testing -- SAT 2015}}, \bibfield{editor}{\bibinfo{person}{M.~Heule} {and}
  \bibinfo{person}{S.~Weaver}} (Eds.). \bibinfo{publisher}{Springer},
  \bibinfo{address}{Cham}, \bibinfo{pages}{104--120}.
\newblock
\showISBNx{978-3-319-24318-4}


\bibitem[Farhi et~al\mbox{.}(2000)]%
        {AQC}
\bibfield{author}{\bibinfo{person}{Edward Farhi}, \bibinfo{person}{Jeffrey
  Goldstone}, \bibinfo{person}{Sam Sam~Gutmann}, {and} \bibinfo{person}{Michael
  Sipser}.} \bibinfo{year}{2000}\natexlab{}.
\newblock \bibinfo{title}{Quantum Computation by Adiabatic Evolution}.
\newblock
\newblock
\urldef\tempurl%
\url{https://doi.org/10.48550/ARXIV.QUANT-PH/0001106}
\showDOI{\tempurl}


\bibitem[Gendreau and Potvin(2005)]%
        {Gendreau05}
\bibfield{author}{\bibinfo{person}{M. Gendreau} {and} \bibinfo{person}{J.
  Potvin}.} \bibinfo{year}{2005}\natexlab{}.
\newblock \bibinfo{booktitle}{\emph{Tabu Search}}.
\newblock \bibinfo{publisher}{Springer}, \bibinfo{address}{Boston, MA},
  \bibinfo{pages}{165--186}.
\newblock
\showISBNx{978-0-387-28356-2}
\urldef\tempurl%
\url{https://doi.org/10.1007/0-387-28356-0_6}
\showDOI{\tempurl}


\bibitem[Gilliam et~al\mbox{.}(2021)]%
        {GroverQUBO}
\bibfield{author}{\bibinfo{person}{Austin Gilliam}, \bibinfo{person}{Stefan
  Woerner}, {and} \bibinfo{person}{Constantin Gonciulea}.}
  \bibinfo{year}{2021}\natexlab{}.
\newblock \showarticletitle{Grover Adaptive Search for Constrained Polynomial
  Binary Optimization}.
\newblock \bibinfo{journal}{\emph{Quantum}}  \bibinfo{volume}{5}
  (\bibinfo{date}{apr} \bibinfo{year}{2021}), \bibinfo{pages}{428}.
\newblock
\urldef\tempurl%
\url{https://doi.org/10.22331/q-2021-04-08-428}
\showDOI{\tempurl}


\bibitem[Glover et~al\mbox{.}(1998)]%
        {Glover98}
\bibfield{author}{\bibinfo{person}{F. Glover}, \bibinfo{person}{G.
  Kochenberger}, {and} \bibinfo{person}{B. Alidaee}.}
  \bibinfo{year}{1998}\natexlab{}.
\newblock \showarticletitle{Adaptive Memory Tabu Search for Binary Quadratic
  Programs}.
\newblock \bibinfo{journal}{\emph{Management Science}}  \bibinfo{volume}{44}
  (\bibinfo{date}{03} \bibinfo{year}{1998}), \bibinfo{pages}{336--345}.
\newblock
\urldef\tempurl%
\url{https://doi.org/10.1287/mnsc.44.3.336}
\showDOI{\tempurl}


\bibitem[Grant et~al\mbox{.}(2021)]%
        {Grant21}
\bibfield{author}{\bibinfo{person}{E. Grant}, \bibinfo{person}{T.~S. Humble},
  {and} \bibinfo{person}{B. Stump}.} \bibinfo{year}{2021}\natexlab{}.
\newblock \showarticletitle{Benchmarking Quantum Annealing Controls with
  Portfolio Optimization}.
\newblock \bibinfo{journal}{\emph{Physical Review Applied}}
  \bibinfo{volume}{15}, \bibinfo{number}{1} (\bibinfo{date}{1}
  \bibinfo{year}{2021}).
\newblock
\showISSN{2331-7019}
\urldef\tempurl%
\url{https://doi.org/10.1103/physrevapplied.15.014012}
\showDOI{\tempurl}


\bibitem[Grover(1996)]%
        {Grover96}
\bibfield{author}{\bibinfo{person}{L.K. Grover}.}
  \bibinfo{year}{1996}\natexlab{}.
\newblock \showarticletitle{A Fast Quantum Mechanical Algorithm for Database
  Search}. In \bibinfo{booktitle}{\emph{Proc. Symposium on Theory of Computing
  (STOC)}} (Philadelphia, Pennsylvania, USA). \bibinfo{publisher}{ACM},
  \bibinfo{address}{New York, NY, USA}, \bibinfo{pages}{212–219}.
\newblock
\showISBNx{0897917855}
\urldef\tempurl%
\url{https://doi.org/10.1145/237814.237866}
\showDOI{\tempurl}


\bibitem[Hassan et~al\mbox{.}(2019)]%
        {HPEC19}
\bibfield{author}{\bibinfo{person}{M.W. Hassan}, \bibinfo{person}{S. Pakin},
  {and} \bibinfo{person}{W. Feng}.} \bibinfo{year}{2019}\natexlab{}.
\newblock \showarticletitle{C to D-Wave: A High-level C Compilation Framework
  for Quantum Annealers}. In \bibinfo{booktitle}{\emph{High Performance Extreme
  Computing Conference (HPEC)}}. \bibinfo{publisher}{IEEE}.
\newblock


\bibitem[Khetawat et~al\mbox{.}(2019)]%
        {Atrey19}
\bibfield{author}{\bibinfo{person}{H. Khetawat}, \bibinfo{person}{A. Atrey},
  \bibinfo{person}{G. Li}, \bibinfo{person}{F. Mueller}, {and}
  \bibinfo{person}{S. Pakin}.} \bibinfo{year}{2019}\natexlab{}.
\newblock \showarticletitle{Implementing NChooseK on IBM Q Quantum Computer
  Systems}. In \bibinfo{booktitle}{\emph{Reversible Computation}},
  \bibfield{editor}{\bibinfo{person}{M.~K. Thomsen} {and}
  \bibinfo{person}{M.~Soeken}} (Eds.). \bibinfo{publisher}{Springer
  International Publishing}, \bibinfo{pages}{209--223}.
\newblock
\urldef\tempurl%
\url{https://doi.org/10.1007/978-3-030-21500-2_13}
\showDOI{\tempurl}


\bibitem[King and McGeoch(2014)]%
        {King14}
\bibfield{author}{\bibinfo{person}{A.~D. King} {and} \bibinfo{person}{C.~C.
  McGeoch}.} \bibinfo{year}{2014}\natexlab{}.
\newblock \bibinfo{title}{Algorithm engineering for a quantum annealing
  platform}.
\newblock
\newblock
\showeprint[arxiv]{1410.2628}~[cs.DS]


\bibitem[King(1976)]%
        {SYMEX}
\bibfield{author}{\bibinfo{person}{J.C. King}.}
  \bibinfo{year}{1976}\natexlab{}.
\newblock \showarticletitle{Symbolic Execution and Program Testing}.
\newblock \bibinfo{journal}{\emph{Commun. ACM}} \bibinfo{volume}{19},
  \bibinfo{number}{7} (\bibinfo{year}{1976}), \bibinfo{pages}{385--394}.
\newblock
\urldef\tempurl%
\url{https://doi.org/10.1145/360248.360252}
\showDOI{\tempurl}


\bibitem[Kirsch(2017)]%
        {Onward17}
\bibfield{author}{\bibinfo{person}{C.M. Kirsch}.}
  \bibinfo{year}{2017}\natexlab{}.
\newblock \showarticletitle{Selfie and the Basics}. In
  \bibinfo{booktitle}{\emph{Proc. ACM SIGPLAN International Symposium on New
  Ideas, New Paradigms, and Reflections on Programming and Software
  (Onward!)}}. \bibinfo{publisher}{ACM}.
\newblock


\bibitem[Kochenberger et~al\mbox{.}(2014)]%
        {Hao14}
\bibfield{author}{\bibinfo{person}{G. Kochenberger}, \bibinfo{person}{J. Hao},
  \bibinfo{person}{F. Glover}, \bibinfo{person}{M. Lewis}, \bibinfo{person}{Z.
  Lü}, \bibinfo{person}{H. Wang}, {and} \bibinfo{person}{Y. Wang}.}
  \bibinfo{year}{2014}\natexlab{}.
\newblock \showarticletitle{The Unconstrained Binary Quadratic Programming
  Problem: A Survey}.
\newblock \bibinfo{journal}{\emph{Journal of Combinatorial Optimization}}
  \bibinfo{volume}{28} (\bibinfo{date}{07} \bibinfo{year}{2014}).
\newblock
\urldef\tempurl%
\url{https://doi.org/10.1007/s10878-014-9734-0}
\showDOI{\tempurl}


\bibitem[Kurowski et~al\mbox{.}(2020)]%
        {Kurowski20}
\bibfield{author}{\bibinfo{person}{K. Kurowski}, \bibinfo{person}{J. Weglarz},
  \bibinfo{person}{M. Subocz}, \bibinfo{person}{R. Rózycki}, {and}
  \bibinfo{person}{G. Waligóra}.} \bibinfo{year}{2020}\natexlab{}.
\newblock \showarticletitle{Hybrid Quantum Annealing Heuristic Method for
  Solving Job Shop Scheduling Problem}. In
  \bibinfo{booktitle}{\emph{Computational Science -- ICCS 2020}},
  \bibfield{editor}{\bibinfo{person}{V.~V. Krzhizhanovskaya},
  \bibinfo{person}{G.~Závodszky}, \bibinfo{person}{M.~H. Lees},
  \bibinfo{person}{J.~J. Dongarra}, \bibinfo{person}{P.~Sloot},
  \bibinfo{person}{S.~Brissos}, {and} \bibinfo{person}{J.~Teixeira}} (Eds.).
  \bibinfo{publisher}{Springer}, \bibinfo{address}{Cham},
  \bibinfo{pages}{502--515}.
\newblock
\showISBNx{978-3-030-50433-5}


\bibitem[Lewis and Glover(2017)]%
        {Lewis17}
\bibfield{author}{\bibinfo{person}{M. Lewis} {and} \bibinfo{person}{F.
  Glover}.} \bibinfo{year}{2017}\natexlab{}.
\newblock \bibinfo{title}{Quadratic Unconstrained Binary Optimization Problem
  Preprocessing: Theory and Empirical Analysis}.
\newblock
\newblock
\showeprint[arxiv]{1705.09844}~[cs.AI]


\bibitem[Lucas(2014)]%
        {Andrew14}
\bibfield{author}{\bibinfo{person}{A. Lucas}.} \bibinfo{year}{2014}\natexlab{}.
\newblock \showarticletitle{Ising formulations of many NP problems}.
\newblock \bibinfo{journal}{\emph{Frontiers in Physics}}  \bibinfo{volume}{2}
  (\bibinfo{year}{2014}).
\newblock
\showISSN{2296-424X}
\urldef\tempurl%
\url{https://doi.org/10.3389/fphy.2014.00005}
\showDOI{\tempurl}


\bibitem[Meza(2010)]%
        {Meza10}
\bibfield{author}{\bibinfo{person}{J.~C. Meza}.}
  \bibinfo{year}{2010}\natexlab{}.
\newblock \showarticletitle{Steepest Descent}.
\newblock \bibinfo{journal}{\emph{WIREs Computational Statistics}}
  \bibinfo{volume}{2}, \bibinfo{number}{6} (\bibinfo{year}{2010}),
  \bibinfo{pages}{719--722}.
\newblock
\urldef\tempurl%
\url{https://doi.org/10.1002/wics.117}
\showDOI{\tempurl}


\bibitem[Misevicius(2005)]%
        {Misevicius05}
\bibfield{author}{\bibinfo{person}{A. Misevicius}.}
  \bibinfo{year}{2005}\natexlab{}.
\newblock \showarticletitle{A Tabu Search Algorithm for the Quadratic
  Assignment Problem}.
\newblock \bibinfo{journal}{\emph{Computational Optimization and Applications}}
   \bibinfo{volume}{30} (\bibinfo{date}{01} \bibinfo{year}{2005}),
  \bibinfo{pages}{95--111}.
\newblock
\urldef\tempurl%
\url{https://doi.org/10.1007/s10589-005-4562-x}
\showDOI{\tempurl}


\bibitem[Nannicini(2019)]%
        {VQA}
\bibfield{author}{\bibinfo{person}{Giacomo Nannicini}.}
  \bibinfo{year}{2019}\natexlab{}.
\newblock \showarticletitle{Performance of hybrid quantum-classical variational
  heuristics for combinatorial optimization}.
\newblock \bibinfo{journal}{\emph{Phys. Rev. E}}  \bibinfo{volume}{99}
  (\bibinfo{date}{Jan} \bibinfo{year}{2019}), \bibinfo{pages}{013304}.
\newblock
Issue 1.
\urldef\tempurl%
\url{https://doi.org/10.1103/PhysRevE.99.013304}
\showDOI{\tempurl}


\bibitem[Niemetz et~al\mbox{.}(2015)]%
        {Boolector}
\bibfield{author}{\bibinfo{person}{A. Niemetz}, \bibinfo{person}{M. Preiner},
  {and} \bibinfo{person}{A. Biere}.} \bibinfo{year}{2015}\natexlab{}.
\newblock \showarticletitle{Boolector 2.0 System Description}.
\newblock \bibinfo{journal}{\emph{Journal on Satisfiability, Boolean Modeling
  and Computation}}  \bibinfo{volume}{9} (\bibinfo{year}{2015}),
  \bibinfo{pages}{53--58}.
\newblock


\bibitem[Niemetz et~al\mbox{.}(2018)]%
        {BTOR2}
\bibfield{author}{\bibinfo{person}{A. Niemetz}, \bibinfo{person}{M. Preiner},
  \bibinfo{person}{C. Wolf}, {and} \bibinfo{person}{A. Biere}.}
  \bibinfo{year}{2018}\natexlab{}.
\newblock \showarticletitle{Btor2, BtorMC and Boolector 3.0}. In
  \bibinfo{booktitle}{\emph{Computer Aided Verification}},
  \bibfield{editor}{\bibinfo{person}{H.~Chockler} {and}
  \bibinfo{person}{G.~Weissenbacher}} (Eds.). \bibinfo{publisher}{Springer},
  \bibinfo{pages}{587--595}.
\newblock


\bibitem[Pakin(2019)]%
        {ASPLOS19}
\bibfield{author}{\bibinfo{person}{S. Pakin}.} \bibinfo{year}{2019}\natexlab{}.
\newblock \showarticletitle{Targeting Classical Code to a Quantum Annealer}. In
  \bibinfo{booktitle}{\emph{Proc. International Conference on Architectural
  Support for Programming Languages and Operating Systems (ASPLOS)}}.
  \bibinfo{publisher}{ACM}.
\newblock
\urldef\tempurl%
\url{https://doi.org/10.1145/3297858.3304071}
\showDOI{\tempurl}


\bibitem[Pakin(2021)]%
        {Pakin21}
\bibfield{author}{\bibinfo{person}{S. Pakin}.} \bibinfo{year}{2021}\natexlab{}.
\newblock \showarticletitle{A Simple Heuristic for Expressing a Truth Table as
  a Quadratic Pseudo-Boolean Function}. In \bibinfo{booktitle}{\emph{2021 IEEE
  International Conference on Quantum Computing and Engineering (QCE)}}.
  \bibinfo{pages}{218--224}.
\newblock
\urldef\tempurl%
\url{https://doi.org/10.1109/QCE52317.2021.00039}
\showDOI{\tempurl}


\bibitem[Paul(2011)]%
        {Gerald11}
\bibfield{author}{\bibinfo{person}{G. Paul}.} \bibinfo{year}{2011}\natexlab{}.
\newblock \showarticletitle{An Efficient Implementation of the Robust Tabu
  Search Heuristic for Sparse Quadratic Assignment Problems}.
\newblock \bibinfo{journal}{\emph{European Journal of Operational Research}}
  \bibinfo{volume}{209} (\bibinfo{date}{03} \bibinfo{year}{2011}),
  \bibinfo{pages}{215--218}.
\newblock
\urldef\tempurl%
\url{https://doi.org/10.1016/j.ejor.2010.09.009}
\showDOI{\tempurl}


\bibitem[Pelofske et~al\mbox{.}(2021)]%
        {Pelofske2021}
\bibfield{author}{\bibinfo{person}{E. Pelofske}, \bibinfo{person}{J. Golden},
  \bibinfo{person}{A. Bärtschi}, \bibinfo{person}{D. O'Malley}, {and}
  \bibinfo{person}{S. Eidenbenz}.} \bibinfo{year}{2021}\natexlab{}.
\newblock \bibinfo{title}{Sampling on NISQ Devices: "Who's the Fairest One of
  All?"}.
\newblock
\newblock
\showeprint[arxiv]{2107.06468}~[quant-ph]


\bibitem[Shylo et~al\mbox{.}(2012)]%
        {Shylo12}
\bibfield{author}{\bibinfo{person}{O. Shylo}, \bibinfo{person}{D. Korenkevych},
  {and} \bibinfo{person}{P. Pardalos}.} \bibinfo{year}{2012}\natexlab{}.
\newblock \showarticletitle{Global Equilibrium Search Algorithms for
  Combinatorial Optimization Problems}. \bibinfo{pages}{277--286}.
\newblock
\showISBNx{978-3-642-32963-0}
\urldef\tempurl%
\url{https://doi.org/10.1007/978-3-642-32964-7_28}
\showDOI{\tempurl}


\bibitem[Shylo and Shylo(2010)]%
        {Shylo10}
\bibfield{author}{\bibinfo{person}{V.~P. Shylo} {and} \bibinfo{person}{O.~V.
  Shylo}.} \bibinfo{year}{2010}\natexlab{}.
\newblock \showarticletitle{Solving the Maxcut Problem by the Global
  Equilibrium Search}.
\newblock \bibinfo{journal}{\emph{Cybernetics and Systems Analysis}}
  \bibinfo{volume}{46}, \bibinfo{number}{5} (\bibinfo{date}{01 9}
  \bibinfo{year}{2010}), \bibinfo{pages}{744--754}.
\newblock
\showISSN{1573-8337}
\urldef\tempurl%
\url{https://doi.org/10.1007/s10559-010-9256-4}
\showDOI{\tempurl}


\bibitem[Su et~al\mbox{.}(2016)]%
        {Su16}
\bibfield{author}{\bibinfo{person}{J. Su}, \bibinfo{person}{T. Tu}, {and}
  \bibinfo{person}{L. He}.} \bibinfo{year}{2016}\natexlab{}.
\newblock \showarticletitle{A Quantum Annealing Approach for Boolean
  Satisfiability Problem}. In \bibinfo{booktitle}{\emph{Proc. ACM/EDAC/IEEE
  Design Automation Conference (DAC)}}. \bibinfo{pages}{1--6}.
\newblock
\urldef\tempurl%
\url{https://doi.org/10.1145/2897937.2897973}
\showDOI{\tempurl}


\bibitem[Tabi et~al\mbox{.}(2020)]%
        {Tabi20}
\bibfield{author}{\bibinfo{person}{Z. Tabi}, \bibinfo{person}{K.~H. El-Safty},
  \bibinfo{person}{Z. Kallus}, \bibinfo{person}{P. Haga}, \bibinfo{person}{T.
  Kozsik}, \bibinfo{person}{A. Glos}, {and} \bibinfo{person}{Z. Zimboras}.}
  \bibinfo{year}{2020}\natexlab{}.
\newblock \showarticletitle{Quantum Optimization for the Graph Coloring Problem
  with Space-Efficient Embedding}.
\newblock \bibinfo{journal}{\emph{Proc. International Conference on Quantum
  Computing and Engineering (QCE)}} (\bibinfo{date}{10} \bibinfo{year}{2020}).
\newblock
\urldef\tempurl%
\url{https://doi.org/10.1109/qce49297.2020.00018}
\showDOI{\tempurl}


\bibitem[Tanburn et~al\mbox{.}(2015)]%
        {Tanburn15}
\bibfield{author}{\bibinfo{person}{R. Tanburn}, \bibinfo{person}{E. Okada},
  {and} \bibinfo{person}{N. Dattani}.} \bibinfo{year}{2015}\natexlab{}.
\newblock \bibinfo{title}{Reducing multi-qubit interactions in adiabatic
  quantum computation without adding auxiliary qubits. Part 1: The
  "deduc-reduc" method and its application to quantum factorization of
  numbers}.
\newblock
\newblock
\showeprint[arxiv]{1508.04816}~[quant-ph]


\bibitem[Tsuchimochi et~al\mbox{.}(2022)]%
        {ImaginaryTimeEvol}
\bibfield{author}{\bibinfo{person}{Takashi Tsuchimochi},
  \bibinfo{person}{Yoohee Ryo}, {and} \bibinfo{person}{Seiichiro~L. Ten-no}.}
  \bibinfo{year}{2022}\natexlab{}.
\newblock \bibinfo{title}{Improved algorithms of quantum imaginary time
  evolution for ground and excited states of molecular systems}.
\newblock
\newblock
\urldef\tempurl%
\url{https://doi.org/10.48550/ARXIV.2205.01983}
\showDOI{\tempurl}


\bibitem[Ushijima-Mwesigwa et~al\mbox{.}(2017)]%
        {Hayato17}
\bibfield{author}{\bibinfo{person}{H. Ushijima-Mwesigwa},
  \bibinfo{person}{C.~F.~A. Negre}, {and} \bibinfo{person}{S.~M. Mniszewski}.}
  \bibinfo{year}{2017}\natexlab{}.
\newblock \bibinfo{title}{Graph Partitioning using Quantum Annealing on the
  D-Wave System}.
\newblock
\newblock
\showeprint[arxiv]{1705.03082}~[quant-ph]


\bibitem[Waidyasooriya and Hariyama(2020)]%
        {Hariyama20}
\bibfield{author}{\bibinfo{person}{H.~M. Waidyasooriya} {and}
  \bibinfo{person}{M. Hariyama}.} \bibinfo{year}{2020}\natexlab{}.
\newblock \showarticletitle{A GPU-Based Quantum Annealing Simulator for
  Fully-Connected Ising Models Utilizing Spatial and Temporal Parallelism}.
\newblock \bibinfo{journal}{\emph{IEEE Access}}  \bibinfo{volume}{8}
  (\bibinfo{year}{2020}), \bibinfo{pages}{67929--67939}.
\newblock
\urldef\tempurl%
\url{https://doi.org/10.1109/ACCESS.2020.2985699}
\showDOI{\tempurl}


\bibitem[Wang et~al\mbox{.}(2020)]%
        {Wang20}
\bibfield{author}{\bibinfo{person}{B. Wang}, \bibinfo{person}{F. Hu},
  \bibinfo{person}{H. Yao}, {and} \bibinfo{person}{C. Wang}.}
  \bibinfo{year}{2020}\natexlab{}.
\newblock \showarticletitle{Prime Factorization Algorithm Based on Parameter
  Optimization of Ising Model}.
\newblock \bibinfo{journal}{\emph{Scientific Reports}}  \bibinfo{volume}{10}
  (\bibinfo{date}{04} \bibinfo{year}{2020}).
\newblock
\urldef\tempurl%
\url{https://doi.org/10.1038/s41598-020-62802-5}
\showDOI{\tempurl}


\bibitem[Waterman et~al\mbox{.}(2016)]%
        {RISCV}
\bibfield{author}{\bibinfo{person}{A. Waterman}, \bibinfo{person}{Y. Lee},
  \bibinfo{person}{D.A. Patterson}, {and} \bibinfo{person}{K. Asanović}.}
  \bibinfo{year}{2016}\natexlab{}.
\newblock \bibinfo{booktitle}{\emph{The RISC-V Instruction Set Manual, Volume
  I: User-Level ISA, Version 2.1}}.
\newblock \bibinfo{type}{{T}echnical {R}eport} UCB/EECS-2016-118.
  \bibinfo{institution}{EECS Department, University of California, Berkeley}.
\newblock
\urldef\tempurl%
\url{http://www2.eecs.berkeley.edu/Pubs/TechRpts/2016/EECS-2016-118.html}
\showURL{%
\tempurl}


\bibitem[Weaver et~al\mbox{.}(2014)]%
        {Weaver14}
\bibfield{author}{\bibinfo{person}{S. Weaver}, \bibinfo{person}{K. Ray},
  \bibinfo{person}{V. Marek}, \bibinfo{person}{A. Mayer}, {and}
  \bibinfo{person}{A. Walker}.} \bibinfo{year}{2014}\natexlab{}.
\newblock \showarticletitle{Satisfiability-based Set Membership Filters}.
\newblock \bibinfo{journal}{\emph{Journal on Satisfiability, Boolean Modeling
  and Computation}}  \bibinfo{volume}{8} (\bibinfo{date}{01}
  \bibinfo{year}{2014}), \bibinfo{pages}{129--148}.
\newblock
\urldef\tempurl%
\url{https://doi.org/10.3233/SAT190095}
\showDOI{\tempurl}


\bibitem[Wilson et~al\mbox{.}(2021)]%
        {Wilson21}
\bibfield{author}{\bibinfo{person}{E. Wilson}, \bibinfo{person}{F. Mueller},
  {and} \bibinfo{person}{S. Pakin}.} \bibinfo{year}{2021}\natexlab{}.
\newblock \showarticletitle{Mapping Constraint Problems onto Quantum Gate and
  Annealing Devices}. In \bibinfo{booktitle}{\emph{2021 IEEE/ACM Second
  International Workshop on Quantum Computing Software (QCS)}}.
  \bibinfo{pages}{110--117}.
\newblock
\urldef\tempurl%
\url{https://doi.org/10.1109/QCS54837.2021.00016}
\showDOI{\tempurl}


\bibitem[Zahedinejad and Zaribafiyan(2017)]%
        {Ehsan17}
\bibfield{author}{\bibinfo{person}{Ehsan Zahedinejad} {and}
  \bibinfo{person}{Arman Zaribafiyan}.} \bibinfo{year}{2017}\natexlab{}.
\newblock \bibinfo{title}{Combinatorial Optimization on Gate Model Quantum
  Computers: A Survey}.
\newblock
\newblock
\urldef\tempurl%
\url{https://doi.org/10.48550/ARXIV.1708.05294}
\showDOI{\tempurl}


\bibitem[Z.Bian et~al\mbox{.}(2014)]%
        {Bian14}
\bibfield{author}{\bibinfo{person}{Z.Bian}, \bibinfo{person}{F. Chudak},
  \bibinfo{person}{R. Israel}, \bibinfo{person}{B. Lackey},
  \bibinfo{person}{W.G. Macready}, {and} \bibinfo{person}{A. Roy}.}
  \bibinfo{year}{2014}\natexlab{}.
\newblock \showarticletitle{Discrete optimization using quantum annealing on
  sparse Ising models}.
\newblock \bibinfo{journal}{\emph{Frontiers in Physics}}  \bibinfo{volume}{2}
  (\bibinfo{year}{2014}).
\newblock
\urldef\tempurl%
\url{https://doi.org/10.3389/fphy.2014.00056}
\showDOI{\tempurl}


\bibitem[Zbinden et~al\mbox{.}(2020)]%
        {Zbinden20}
\bibfield{author}{\bibinfo{person}{S. Zbinden}, \bibinfo{person}{A.
  B{\"a}rtschi}, \bibinfo{person}{H. Djidjev}, {and} \bibinfo{person}{S.
  Eidenbenz}.} \bibinfo{year}{2020}\natexlab{}.
\newblock \showarticletitle{Embedding Algorithms for Quantum Annealers with
  Chimera and Pegasus Connection Topologies}. In \bibinfo{booktitle}{\emph{High
  Performance Computing}}, \bibfield{editor}{\bibinfo{person}{P.~Sadayappan},
  \bibinfo{person}{B.~Chamberlain}, \bibinfo{person}{G.~Juckeland}, {and}
  \bibinfo{person}{H.~Ltaief}} (Eds.). \bibinfo{publisher}{Springer},
  \bibinfo{pages}{187--206}.
\newblock
\showISBNx{978-3-030-50743-5}


\bibitem[Zhou et~al\mbox{.}(2013)]%
        {Zhou13}
\bibfield{author}{\bibinfo{person}{Y. Zhou}, \bibinfo{person}{J. Wang}, {and}
  \bibinfo{person}{J. Yin}.} \bibinfo{year}{2013}\natexlab{}.
\newblock \showarticletitle{A Directional-Biased Tabu Search algorithm for
  Multi-Objective Unconstrained Binary Quadratic Programming Problem}.
\newblock \bibinfo{journal}{\emph{Proc. International Conference on Advanced
  Computational Intelligence (ICACI)}} (\bibinfo{year}{2013}),
  \bibinfo{pages}{281--286}.
\newblock


\end{thebibliography}

\newpage

\section*{Supplementary Material}

In the following, we present the BEATOR algorithm in detail and an outline of the QUBOT and QUARC algorithms to support the propositions in the paper. Formal proofs remain future work which is likely to involve proof assistants and automation.

Given a C* program $P$ compiled to a $w$-bit RISC-U binary $R$, BEATOR translates $R$ in $\mathcal{O}(m\cdot|P|)$ time to a $\mathcal{O}(m\cdot|P|)$ BTOR2 model $B$ where $w$ is either $64$ or $32$ and $m$ is either $1$, for all BEATOR configurations other than MMURAM and RAM, or a bound on the size of memory accessed by $P$. QUBOT translates $B$ in $\mathcal{O}(n\cdot m\cdot w^2\cdot|P|)$ time to a $\mathcal{O}(n\cdot m\cdot w^2\cdot|P|)$ QUBO model $Q$ where $n$ is a bound on the number of executed RISC-U instructions in $R$. Since every RISC-U instruction may only increase the size of memory by at most one memory word, $n$ and $m$ may replace each other. Finally, QUARC translates $B$ in $\mathcal{O}(n\cdot m\cdot w\cdot|P|)$ time to a $\mathcal{O}(n\cdot m\cdot w\cdot|P|)$ quantum circuit $C$.

The BEATOR algorithm shown below is for the 32-bit $\mathcal{O}(m\cdot|P|)$ MMURAM configuration which models memory by a finite number of 32-bit memory words avoiding the use of the theory of bitvector arrays through \texttt{read} and \texttt{write} operators. BEATOR executes the pseudo code in the order in which it appears here. Lines that contain a BTOR2 keyword generate output. The nid of each generated line is either represented by a constant or an identifier that starts with \texttt{nid\_}. Variables are printed in italics. Code that checks for errors in the input is omitted. The full BEATOR algorithm for all configurations is available in source code at \url{https://github.com/cksystemsteaching/selfie/blob/unicorn/tools/beator.c}.

The QUBOT and QUARC algorithms shown below are for BTOR2 models generated with the MMURAM or RAM configurations of BEATOR. They use bit blasting to implement all combinational BTOR2 circuits. We nevertheless only show the algorithmic complexity in number of qubits (binary variables). The full QUBOT and QUARC algorithms are available in source code at \url{https://github.com/cksystemsteaching/selfie/tree/unicorn/tools/qubot} and \url{https://github.com/cksystemsteaching/selfie/tree/unicorn/tools/quarc}, respectively.

\footnotesize

\begin{figure}
\centering
\begin{lstlisting}[mathescape,morekeywords={sort,bitvec,zero,one,constd,state,init,not,input,uext}]
1 sort bitvec 1 ; Boolean
2 sort bitvec 32 ; 32-bit machine word
6 sort bitvec $physical\_address\_space\_size\_in\_bits$ ; example 4

10 zero 1
11 one 1

20 zero 2
21 one 2
22 constd 2 2
23 constd 2 3
24 constd 2 4

; start of data segment in 32-bit virtual memory
30 constd 2 $start\_of\_data\_segment$ ; example: 0x11000
; end of data segment in 32-bit virtual memory
31 constd 2 $end\_of\_data\_segment$ ; example: 0x11008

; start of heap segment in 32-bit virtual memory (initial program break)
32 constd 2 $start\_of\_heap\_segment$ ; example: 0x12000
; current end of heap segment in 32-bit virtual memory (current program break)
33 constd 2 $end\_of\_heap\_segment$ ; example: 0x12004

; allowed end of heap segment in 32-bit virtual memory (with 0B allowance)
34 constd 2 $allowed\_end\_of\_heap\_segment$ ; example: 0x12004
; allowed start of stack segment in 32-bit virtual memory (with 0B allowance)
35 constd 2 $allowed\_start\_of\_stack\_segment$ ; example: 0xFFFFFFDC

; highest address in 32-bit virtual address space (4GB)
50 constd 2 4294967292 ; 0xFFFFFFFC

; kernel-mode flag
60 state 1 kernel-mode
61 init 1 60 10 kernel-mode ; initial value is false
62 not 1 60

; unsigned-extended inputs for byte-wise reading
71 sort bitvec 8 ; 1 byte
72 sort bitvec 16 ; 2 bytes
73 sort bitvec 24 ; 3 bytes

81 input 71 1-byte-input
82 input 72 2-byte-input
83 input 73 3-byte-input

91 uext 2 81 24 ; uext-1-byte-input
92 uext 2 82 16 ; uext-2-byte-input
93 uext 2 83 8 ; uext-3-byte-input
94 input 2 4-byte-input
\end{lstlisting}
\caption{32-bit BEATOR header}
\label{fig:beator-header}
\end{figure}

\begin{figure}
\centering
\begin{lstlisting}[mathescape,morekeywords={constd,state,init}]
; 32 32-bit general-purpose registers
; non-zero initial register values
for ($r$ = 1; $r$ < 32; $r$++)
  if ($register\_value[r]$ != 0)
    nid_$r$_initial_value constd 2 $register\_value[r]$

; registers
nid_$0$_register zero 2 zero ; register 0 is always 0
for ($r$ = 1; $r$ < 32; $r$++) {
  nid_$r$_register state 2 $register\_name[r]$
  $nid\_register\_flow[r]$ = nid_$r$_register;
}

; initializing registers
for ($r$ = 1; $r$ < 32; $r$++)
  if ($register\_value[r]$ != 0)
    nid_$r$_initialize_register init 2 nid_$r$_register nid_$r$_initial_value $register\_name[r]$
\end{lstlisting}
\caption{32-bit BEATOR register definitions}
\label{fig:beator-registers}
\end{figure}

\begin{figure}
\centering
\begin{lstlisting}[mathescape,morekeywords={state,init}]
; 32-bit program counter encoded in Boolean flags
for ($pc$ = $start\_of\_code\_segment$; $pc$ < $end\_of\_code\_segment$; $pc$ = $pc$ + 4)
  if ($pc$ is reachable) {
    nid_$pc$_flag: state 1
    if ($pc$ == $entry\_point$)
      nid_$pc$_flag_initial_value: init 1 nid_$pc$_flag 11 ; initial program counter
    else
      nid_$pc$_flag_initial_value: init 1 nid_$pc$_flag 10
  }
\end{lstlisting}
\caption{32-bit BEATOR pc flag definitions}
\label{fig:beator-pc-flags}
\end{figure}

\begin{figure}
\centering
\begin{lstlisting}[mathescape,morekeywords={constd,state,init}]
; physical memory
; data segment
for ($v$ = $start\_of\_data\_segment$, $p$ = 0; $v$ <= $highest\_32\_bit\_address$; $v$ = $v$ + 4, $p$ = $p$ + 1) {
  if ($v$ == $end\_of\_data\_segment$) {
    ; heap segment
    $v$ = $start\_of\_heap\_segment$;
  }
  if ($v$ == $allowed\_end\_of\_heap\_segment$) {
    ; stack segment
    $v$ = $allowed\_start\_of\_stack\_segment$;
  }
  nid_$p$_virtual_address   constd 2 $v$ ; example: 0x11000 vaddr
  nid_$p$_initial_value     constd 2 $memory\_value[v]$ ; example: 0x12000 word
  nid_$p$_RAM_word          state 2 RAM-word-$p$
  nid_$p$_initialize_memory init 2 nid_$p$_RAM_word nid_$p$_initial_value
  $nid\_RAM\_write\_flow[p]$ = nid_$p$_RAM_word;
}
$size\_of\_physical\_memory\_in\_words$ = $p$
\end{lstlisting}
\caption{32-bit BEATOR physical memory}
\label{fig:beator-physical-memory}
\end{figure}

\begin{figure}
\centering
\begin{lstlisting}[mathescape,morekeywords={constd,ite,add,ult,uext,add,sub,mul,udiv,urem}]
; data flow
$nid\_division\_flow$      = 21; // division by one is ok
$nid\_remainder\_flow$     = 21; // remainder by one is ok
$nid\_start\_access\_flow$ = 30; // access in data segment is ok
$nid\_ecall\_flow$         = 10; // initially, no ecall active
for ($pc$ = $start\_of\_code\_segment$; $pc$ < $end\_of\_code\_segment$; $pc$ = $pc$ + 4) {
  if ($pc$: lui rd,imm) {
    if (rd != zero) {
      nid_$pc$_df0 constd 2 imm<<12
      nid_$pc$_df1 ite 2 nid_$pc$_flag nid_$pc$_df0 $nid\_register\_flow[rd]$
      $nid\_register\_flow[rd]$ = nid_$pc$_df1;
    }
  } else if ($pc$: addi rd,rs1,imm) {
    if (rd != zero) {
      nid_$pc$_df0 constd 2 imm
      nid_$pc$_df1 add 2 nid_$rs1$_register nid_$pc$_df0
      nid_$pc$_df2 ite 2 nid_$pc$_flag nid_$pc$_df1 $nid\_register\_flow[rd]$
      $nid\_register\_flow[rd]$ = nid_$pc$_df2;
    }
  } else if ($pc$: [add|sub|mul|divu|remu|sltu] rd,rs1,rs2) {
    if (rd != zero) {
      if ($pc$: divu rd,rs1,rs2) {
        nid_$pc$_df0 ite 2 nid_$pc$_flag nid_$rs2$_register $nid\_division\_flow$
        $nid\_division\_flow$ = nid_$pc$_df0;
      } else if ($pc$: remu rd,rs1,rs2) {
        nid_$pc$_df0 ite 2 nid_$pc$_flag nid_$rs2$_register $nid\_remainder\_flow$
        $nid\_remainder\_flow$ = nid_$pc$_df0;
      }
      if ($pc$: sltu rd,rs1,rs2) {
        nid_$pc$_df0 ult 1 nid_$rs1$_register nid_$rs2$_register
        nin_$pc$_df1 uext 2 nid_$pc$_df0 31
      } else
        nid_$pc$_df1 [add|sub|mul|udiv|urem] 2 nid_$rs1$_register nid_$rs2$_register
      nid_$pc$_df2 ite 2 nid_$pc$_flag nid_$pc$_df1 $nid\_register\_flow[rd]$
      $nid\_register\_flow[rd]$ = nid_$pc$_df2;
    }
  } else ...
\end{lstlisting}
\caption{32-bit BEATOR data flow part I}
\label{fig:beator-data-flow-1}
\end{figure}

\begin{figure}
\centering
\begin{lstlisting}[mathescape,morekeywords={constd,add,ite,eq,not}]
  ... if ($pc$: lw rd,rs1,imm) {
    if (rd != zero) {
      nid_$pc$_lw_imm constd 2 imm
      nid_$pc$_lw_add add 2 nid_$rs1$_register nid_$pc$_lw_imm
      nid_$pc$_lw_acc ite 2 nid_$pc$_flag nid_$pc$_lw_add $nid\_start\_access\_flow$
      $nid\_start\_access\_flow$ = nid_$pc$_lw_acc;
      $nid\_RAM\_read\_flow$ = 20; // segmentation fault checks prevent access
      for ($p$ = 0; $p$ < $size\_of\_physical\_memory\_in\_words$; $p$ = $p$ + 1) {
        nid_$pc$_lw_at_$p$   eq 1 nid_$pc$_lw_add nid_$p$_virtual_address
        nid_$pc$_lw_from_$p$ ite 2 nid_$pc$_lw_at_$p$ nid_$p$_RAM_word $nid\_RAM\_read\_flow$
        $nid\_RAM\_read\_flow$ = nid_$pc$_lw_from_$p$;
      }
      nid_$pc$_lw_val ite 2 nid_$pc$_flag $nid\_RAM\_read\_flow$ $nid\_register\_flow[rd]$
      $nid\_register\_flow[rd]$ = nid_$pc$_lw_val;
    }
  } else if ($pc$: sw rs1,rs2,imm) {
    nid_$pc$_sw_imm constd 2 imm
    nid_$pc$_sw_add add 2 nid_$rs1$_register nid_$pc$_sw_imm
    nid_$pc$_sw_acc ite 2 nid_$pc$_flag nid_$pc$_sw_add $nid\_start\_access\_flow$
    $nid\_start\_access\_flow$ = nid_$pc$_sw_acc;
    for ($p$ = 0; $p$ < $size\_of\_physical\_memory\_in\_words$; $p$ = $p$ + 1) {
      nid_$pc$_sw_at_$p$   eq 1 nid_$pc$_sw_add nid_$p$_virtual_address
      nid_$pc$_sw_to_$p$   ite 2 nid_$pc$_sw_at_$p$ nid_$rs2$_register $nid\_RAM\_write\_flow[p]$
      nid_$pc$_sw_into_$p$ ite 2 nid_$pc$_flag nid_$pc$_sw_to_$p$ $nid\_RAM\_write\_flow[p]$
      $nid\_RAM\_write\_flow[p]$ = nid_$pc$_sw_into_$p$;
    }
  } else if ($pc$: beq rs1,rs2,imm) {
    nid_$pc$_beq eq 1 nid_$rs1$_register nid_$rs2$_register
    nid_$pc$_neq not 1 nid_$pc$_beq
  } else if ($pc$: jal rd,imm) {
    if (rd != zero) {
      nid_$pc$_link constd 2 $pc+4$
      nid_$pc$_ret  ite 2 nid_$pc$_flag nid_$pc$_link $nid\_register\_flow[rd]$
      $nid\_register\_flow[rd]$ = nid_$pc$_ret;
    }
  } else if ($pc$: jalr rd,rs1,imm) {
    // rd update unsupported
  } else if ($pc$: ecall) {
    nid_$pc$_df0 ite 1 nid_$pc$_flag 11 $nid\_ecall\_flow$
    $nid\_ecall\_flow$ = nid_$pc$_df0;
  }
}
\end{lstlisting}
\caption{32-bit BEATOR data flow part II}
\label{fig:beator-data-flow-2}
\end{figure}

\begin{figure}
\centering
\begin{lstlisting}[mathescape]
$current\_callee$ = $start\_of\_code\_segment$;
for ($pc$ = $start\_of\_code\_segment$; $pc$ < $end\_of\_code\_segment$; $pc$ = $pc$ + 4) {
  if ($pc$: [lui|add|sub|mul|divu|remu|sltu|lw|sw])
    $control\_in[pc+4]$ =
      new control_out([lui|add|sub|mul|divu|remu|sltu|lw|sw], $pc$, 0, $control\_in[pc+4]$);
  else if ($pc$: addi rd,rs1,imm) {
    if (rs1 == zero)
      if (imm != 0)
        if (rd == a7)
          // assert: next instruction is ecall
          $reg\_a7$ = imm;
    $control\_in[pc+4]$ = new control_out(addi, $pc$, 0, $control\_in[pc+4]$);
  } else if ($pc$: beq rs1,rs2,imm) {
    $control\_in[pc+imm]$ = new control_out(beq, $pc$, nid_$pc$_beq, $control\_in[pc+imm]$);
    $control\_in[pc+4]$   = new control_out(beq, $pc$, nid_$pc$_neq, $control\_in[pc+4]$);
  } else if ($pc$: jal rd,imm) {
    $control\_in[pc+imm]$ = new control_out(jal, $pc$, 0, $control\_in[pc+imm]$);
    if (rd != zero)
      $control\_in[pc+4]$ =
        new control_out(jalr, $pc+imm$, nid_$pc$_link, $control\_in[pc+4]$);
  } else if ($pc$: jalr rd,rs1,imm) {
    if (rd == zero)
      if (imm == 0)
        if (rs1 == ra) {
          // assert: current callee points to an instruction to which a jal jumps
          $call\_return[current\_callee]$ = $pc$;
          // assert: next "procedure body" begins right after jalr
          $current\_callee$ = $pc+4$;
        }
  } else if ($pc$: ecall) {
    if ($reg\_a7$ == SYSCALL_EXIT)
      // assert: exit ecall is immediately followed by first "procedure body" in code
      $current\_callee$ = $pc+4$;
    $reg\_a7$ = 0;
    $control\_in[pc+4]$ = new control_out(ecall, $pc$, 0, $control\_in[pc+4]$);
  }
}
\end{lstlisting}
\caption{32-bit BEATOR control flow preparation}
\label{fig:beator-control-flow-prep}
\end{figure}

\begin{figure}
\centering
\begin{lstlisting}[mathescape,morekeywords={constd,eq,and,ite}]
; syscalls
nid_syscall_id_exit   constd 2 93 ; SYSCALL_EXIT
nid_syscall_id_read   constd 2 63 ; SYSCALL_READ
nid_syscall_id_write  constd 2 64 ; SYSCALL_WRITE
nid_syscall_id_openat constd 2 56 ; SYSCALL_OPENAT
nid_syscall_id_brk    constd 2 214 ; SYSCALL_BRK

nid_exit_syscall   eq 1 nid_$a7$_register nid_syscall_id_exit ; a7 == SYSCALL_EXIT
nid_read_syscall   eq 1 nid_$a7$_register nid_syscall_id_read ; a7 == SYSCALL_READ
nid_write_syscall  eq 1 nid_$a7$_register nid_syscall_id_write ; a7 == SYSCALL_WRITE
nid_openat_syscall eq 1 nid_$a7$_register nid_syscall_id_openat ; a7 == SYSCALL_OPENAT
nid_brk_syscall    eq 1 nid_$a7$_register nid_syscall_id_brk ; a7 == SYSCALL_BRK

nid_exit_active and 1 $nid\_ecall\_flow$ nid_exit_syscall ; exit ecall is active
; stay in kernel mode indefinitely if exit ecall is active
nid_exit_kernel ite 1 60 nid_exit_syscall nid_exit_active
$nid\_kernel\_mode\_flow$ = nid_exit_kernel;
\end{lstlisting}
\caption{32-bit BEATOR system call header and \texttt{exit} system call}
\label{fig:beator-exit-syscall}
\end{figure}

\begin{figure}
\centering
\begin{lstlisting}[mathescape,morekeywords={and,ite,sub,ugte,eq,add,ult,ugt}]
nid_read_active and 1 $nid\_ecall\_flow$ nid_read_syscall ; read ecall is active
; a1 is start address of buffer for checking address validity
nid_read_access ite 2 nid_read_active nid_$a1$_register $nid\_start\_access\_flow$
$nid\_start\_access\_flow$ = nid_read_access;

; go into kernel mode if read ecall is active
nid_read_kernel ite 1 nid_read_active 11 $nid\_kernel\_mode\_flow$
$nid\_kernel\_mode\_flow$ = nid_read_kernel;

; set a0 = 0 bytes if read ecall is active
nid_read_set_a0 ite 2 nid_read_active 20 $nid\_register\_flow\{a0\}$
$nid\_register\_flow\{a0\}$ = nid_read_set_a0;

nid_read_sub    sub 2 nid_$a2$_register nid_$a0$_register ; a2 - a0
nid_read_cmp    ugte 1 nid_read_sub 24 ; a2 - a0 >= 4 bytes
; read 4 bytes if a2 - a0 >= 4 bytes, or else a2 - a0 bytes
nid_read_inc    ite 2 nid_read_cmp 24 nid_read_sub
nid_read_inc2   eq 1 nid_read_inc 22 ; increment == 2
; unsigned-extended 2-byte input if increment == 2, or else unsigned-extended 1-byte input
nid_read_input2 ite 2 nid_read_inc2 92 91
nid_read_inc3   eq 1 nid_read_inc 23 ; increment == 3
; unsigned-extended 3-byte input if increment == 3
nid_read_input3 ite 2 nid_read_inc3 93 nid_read_input2
nid_read_inc4   eq 1 nid_read_inc 24 ; increment == 4
nid_read_input4 ite 2 nid_read_inc4 94 nid_read_input3 ; 4-byte input if increment == 4
nid_read_cursor add 2 nid_$a1$_register nid_$a0$_register ; a1 + a0
nid_read_more   ult 1 nid_$a0$_register nid_$a2$_register ; a0 < a2
nid_read_goon   and 1 nid_read_syscall nid_read_more ; a7 == SYSCALL_READ and a0 < a2

nid_read_active_kernel and 1 60 nid_read_goon ; read ecall is in kernel mode and not done yet
nid_read_ugti          ugt 1 nid_read_inc 20 ; increment > 0
; read ecall is in kernel mode and not done yet and increment > 0
nid_read_kernel_inc    and 1 nid_read_active_kernel nid_read_ugti

for ($p$ = 0; $p$ < $size\_of\_physical\_memory\_in\_words$; $p$ = $p$ + 1) {
  nid_read_at_$p$     eq 1 nid_read_cursor nid_$p$_virtual_address
  nid_read_input_$p$ ite 2 nid_read_at_$p$ nid_read_input4 $nid\_RAM\_write\_flow[p]$
  ; read input into RAM[a1 + a0]
  nid_read_into_$p$   ite 2 nid_read_kernel_inc nid_read_input_$p$ $nid\_RAM\_write\_flow[p]$
  $nid\_RAM\_write\_flow[p]$ = nid_read_into_$p$;
}

nid_read_move_cursor add 2 nid_$a0$_register nid_read_inc ; a0 + increment
; set a0 = a0 + increment if read ecall is in kernel mode and not done yet
nid_read_set_cursor  ite 2 nid_read_active_kernel nid_read_move_cursor $nid\_register\_flow\{a0\}$
$nid\_register\_flow\{a0\}$ = nid_read_set_cursor;

; stay in kernel mode if read ecall is in kernel mode and not done yet
nid_read_stay ite 1 nid_read_active_kernel 11 $nid\_kernel\_mode\_flow$
$nid\_kernel\_mode\_flow$ = nid_read_stay;
\end{lstlisting}
\caption{32-bit BEATOR \texttt{read} system call}
\label{fig:beator-read-syscall}
\end{figure}

\begin{figure}
\centering
\begin{lstlisting}[mathescape,morekeywords={and,ite,state,init,inc,next,ulte,ult,eq,not}]
nid_write_active and 1 $nid\_ecall\_flow$ nid_write_syscall ; write ecall is active
; a1 is start address of buffer for checking address validity
nid_write_access ite 2 nid_write_active 211 $nid\_start\_access\_flow$
$nid\_start\_access\_flow$ = nid_write_access;

; set a0 = a2 if write ecall is active
nid_write_set_a0 ite 2 nid_write_active nid_$a2$_register $nid\_register\_flow\{a0\}$
$nid\_register\_flow\{a0\}$ = nid_write_set_a0;


nid_openat_active and 1 $nid\_ecall\_flow$ nid_openat_syscall ; openat ecall is active
; a1 is start address of buffer for checking address validity
nid_openat_access ite 2 nid_openat_active nid_$a1$_register $nid\_start\_access\_flow$
$nid\_start\_access\_flow$ = nid_openat_access;

nid_openat_bump   state 2 fd-bump-pointer
; initial fd-bump-pointer is 1 (file descriptor bump pointer)
nid_openat_init   init 2 nid_openat_bump 21
nid_openat_inc    inc 2 nid_openat_bump
; fd-bump-pointer + 1 if openat ecall is active
nid_openat_newfd  ite 2 nid_openat_active nid_fd_inc nid_fd_bump
; increment fd-bump-pointer if openat ecall is active
nid_openat_next   next 2 nid_fd_bump nid_openat_newfd
; set a0 = fd-bump-pointer + 1 if openat ecall is active
nid_openat_set_a0 ite 2 nid_openat_active nid_fd_inc $nid\_register\_flow\{a0\}$
$nid\_register\_flow\{a0\}$ = nid_openat_set_a0;


nid_brk_active  and 1 $nid\_ecall\_flow$ nid_brk_syscall ; brk ecall is active
nid_brk_bump    state 2 brk-bump-pointer
nid_brk_init    init 2 nid_brk_bump 33 ; current program break
nid_brk_shrink  ulte 1 nid_brk_bump nid_$a0$_register ; brk <= a0
nid_brk_free    ult 1 nid_$a0$_register nid_$sp$_register ; a0 < sp
nid_brk_segment and 1 nid_brk_shrink nid_brk_free ; brk <= a0 < sp
nid_brk_reset   and 2 nid_$a0$_register 23 ; reset all but 2 LSBs of a0
nid_brk_aligned eq 1 nid_brk_reset 20 ; 2 LSBs of a0 == 0 (a0 is word-aligned)
; brk <= a0 < sp and a0 is word-aligned (a0 is valid)
nid_brk_valid   and 1 nid_brk_segment nid_brk_aligned
nid_brk_ok      and 1 nid_brk_active nid_brk_valid ; brk ecall is active and a0 is valid
; brk = a0 if brk ecall is active and a0 is valid
nid_brk_newbrk  ite 2 nid_brk_ok nid_$a0$_register nid_brk_bump
; set brk = a0 if brk ecall is active and a0 is valid
nid_brk_next    next 2 nid_brk_bump nid_brk_newbrk
nid_brk_invalid not 1 nid_brk_valid ; a0 is invalid
nid_brk_fail    and 1 nid_brk_active nid_brk_invalid ; brk ecall is active and a0 is invalid
; set a0 = brk if brk ecall is active and a0 is invalid
nid_brk_set_a0 ite 2 nid_brk_fail nid_brk_bump $nid\_register\_flow\{a0\}$
$nid\_register\_flow\{a0\}$ = nid_brk_set_a0;

nid_update_kernel next 1 60 $nid\_kernel\_mode\_flow$ ; updating kernel-mode flag
\end{lstlisting}
\caption{32-bit BEATOR \texttt{write}, \texttt{openat}, and \texttt{brk} system calls}
\label{fig:beator-remaining-syscalls}
\end{figure}

\begin{figure}
\centering
\begin{lstlisting}[mathescape,morekeywords={and,not,eq,state,init,ite,next}]
; control flow
for ($pc$ = $start\_of\_code\_segment$; $pc$ < $end\_of\_code\_segment$; $pc$ = $pc$ + 4) {
  $nid\_control\_flow$ = 10;
  $control\_out$       = $control\_in[pc]$;
  while ($control\_out$ != 0) {
    ($from\_is$, $from\_pc$, $nid\_condition$, $control\_out$) = $control\_out$;
    if ($from\_pc$ != 0) {
      if ($from\_is$ == beq) {
        nid_$pc$_active and 1 nid_$from\_pc$_flag $nid\_condition$
        $nid\_from\_active$ = nid_$pc$_active;
      } else if ($from\_is$ == jalr) {
        $from\_pc$ = $call\_return[from\_pc]$;
        nid_$pc$_not    not 2 21
        nid_$pc$_lsbrst and 2 nid_$ra$_register nid_$pc$_not
        nid_$pc$_equal  eq 1 nid_$pc$_lsbrst $nid\_condition$
        nid_$pc$_active and 1 nid_$from\_pc$_flag nid_$pc$_equal
        $nid\_from\_active$ = nid_$pc$_active;
      } else if ($from\_is$ == ecall) {
        nid_$pc$_kernel   state 1 kernel-mode-pc-flag-$from\_pc$
        nid_$pc$_inactive init 1 nid_$pc$_kernel 10 ; ecall is initially inactive
        ; activate ecall and keep active while in kernel mode
        nid_$pc$_active   ite 1 nid_$pc$_kernel 60 nid_$from\_pc$_flag
        ; keep ecall active while in kernel mode
        nid_$pc$_stay     next 1 nid_$pc$_kernel nid_$pc$_active
        ; ecall is active but not in kernel mode anymore
        nid_$pc$_leave    and 1 nid_$pc$_kernel 62
        $nid\_from\_active$ = nid_$pc$_leave;
      } else
        $nid\_from\_active$ = nid_$from\_pc$_flag;
      if ($nid\_control\_flow$ == 10)
        $nid\_control\_flow$ = $nid\_from\_active$;
      else {
        nid_$pc$_next ite 1 $nid\_from\_active$ 11 $nid\_control\_flow$
        $nid\_control\_flow$ = nid_$pc$_next;
      }
    }
  }
  nid_$pc$_next next 1 nid_$pc$_flag $nid\_control\_flow$
}
\end{lstlisting}
\caption{32-bit BEATOR control flow}
\label{fig:beator-control-flow}
\end{figure}

\begin{figure}
\centering
\begin{lstlisting}[mathescape,morekeywords={next}]
; updating registers
for ($r$ = 1; $r$ < 32; $r$++)
  nid_$r$_register_update next 2 nid_$r$_register $nid\_register\_flow[r]$ $register\_name[r]$

; updating physical memory
for ($p$ = 0; $p$ < $size\_of\_physical\_memory\_in\_words$; $p$ = $p$ + 1)
  nid_$p$_memory_update next 2 nid_$p$_RAM_word $nid\_RAM\_write\_flow[p]$ RAM-word-$p$
\end{lstlisting}
\caption{32-bit BEATOR register and memory update}
\label{fig:beator-update}
\end{figure}

\begin{figure}
\centering
\begin{lstlisting}[mathescape,morekeywords={not,and,bad,neq,eq}]
; checking syscall id
nid_check_syscall0 not 1 nid_syscall_id_exit ; a7 != SYSCALL_EXIT
nid_check_syscall1 not 1 nid_syscall_id_read ; a7 != SYSCALL_READ
nid_check_syscall2 not 1 nid_syscall_id_write ; a7 != SYSCALL_WRITE
nid_check_syscall3 not 1 nid_syscall_id_openat ; a7 != SYSCALL_OPENAT
nid_check_syscall4 not 1 nid_syscall_id_brk ; a7 != SYSCALL_BRK
nid_check_syscall5 and 1 nid_check_syscall0 nid_check_syscall1 ; & a7 != SYSCALL_READ
nid_check_syscall6 and 1 nid_check_syscall5 nid_check_syscall2 ; & a7 != SYSCALL_WRITE
nid_check_syscall7 and 1 nid_check_syscall6 nid_check_syscall3 ; & a7 != SYSCALL_OPENAT
nid_check_syscall8 and 1 nid_check_syscall7 nid_check_syscall4 ; & a7 != SYSCALL_BRK
; ecall is active for invalid syscall id
nid_check_syscall9 and 1 $nid\_ecall\_flow$ nid_check_syscall8
nid_check_syscall10 bad nid_check_syscall9 b0 ; invalid syscall id

; checking exit code
nid_check_exit_code0 neq 1 nid-$a0$-register 20 ; a0 != zero exit code
; exit ecall is active with non-zero exit code
nid_check_exit_code1 and 1 nid_exit_active nid_check_exit_code0
nid_check_exit_code2 bad nid_check_exit_code1 b1 ; non-zero exit code

; checking division and remainder by zero
nid_check_division0  eq 1 $nid\_division\_flow$ 20
nid_check_division1  bad nid_check_division0 b2 ; division by zero
nid_check_remainder0 eq 1 $nid\_remainder\_flow$ 20
nid_check_remainder1 bad nid_check_remainder0 b3 ; remainder by zero

; checking address validity
; is start address of memory access word-aligned?
nid_check_address0 and 2 $nid\_start\_access\_flow$ 23 ; reset all but 2 LSBs of address
; 2 LSBs of address != 0 (address is not word-aligned)
nid_check_address1 neq 1 nid_check_address0 20
nid_check_address2 bad nid_check_address1 b4 ; word-unaligned memory access
\end{lstlisting}
\caption{32-bit BEATOR bad states other than segmentation faults}
\label{fig:beator-bad-states}
\end{figure}

\begin{figure}
\centering
\begin{lstlisting}[mathescape,morekeywords={ult,bad,ugte,ult,and,ugt}]
; checking segmentation faults
; is start address of memory access in a valid segment?
nid_check_segfault0  ult 1 $nid\_start\_access\_flow$ 30 ; address < start of data segment
nid_check_segfault1  bad nid_check_segfault0 b6 ; memory access below data segment
nid_check_segfault2  ugte 1 $nid\_start\_access\_flow$ 31 ; address >= end of data segment
nid_check_segfault3  ult 1 $nid\_start\_access\_flow$ 32 ; address < start of heap segment
nid_check_segfault4  and 1 nid_check_segfault2 nid_check_segfault3
; memory access in between data and heap segments
nid_check_segfault5  bad nid_check_segfault4 b7
; address >= current end of heap segment
nid_check_segfault6  ugte 1 $nid\_start\_access\_flow$ nid_bump_pointer
; address < current start of stack segment
nid_check_segfault7  ult 1 $nid\_start\_access\_flow$ nid-$sp$-register
nid_check_segfault8  and 1 nid_check_segfault6 nid_check_segfault7
; memory access in between current heap and stack segments
nid_check_segfault9  bad nid_check_segfault8 b8
; address >= allowed end of heap segment
nid_check_segfault10 ugte 1 $nid\_start\_access\_flow$ 34
; address < current end of heap segment
nid_check_segfault11 ult 1 $nid\_start\_access\_flow$ nid_bump_pointer
nid_check_segfault12 and 1 nid_check_segfault10 nid_check_segfault11
; memory access in between allowed and current end of heap segment
nid_check_segfault13 bad nid_check_segfault12 b9
; address >= current start of stack segment
nid_check_segfault14 ugte 1 $nid\_start\_access\_flow$ nid-$sp$-register
; address < allowed start of stack segment
nid_check_segfault15 ult 1 $nid\_start\_access\_flow$ 35
nid_check_segfault16 and 1 nid_check_segfault14 nid_check_segfault15
; memory access in between current and allowed start of stack segment
nid_check_segfault17 bad nid_check_segfault16 b10
; address > highest address in 32-bit virtual address space (4GB)
nid_check_segfault18 ugt 1 $nid\_start\_access\_flow$ 50
; memory access above stack segment
nid_check_segfault19 bad nid_check_segfault18 b11
\end{lstlisting}
\caption{32-bit BEATOR segmentation fault checks}
\label{fig:beator-segfaults}
\end{figure}

\begin{figure}
\centering
\begin{lstlisting}[mathescape,morekeywords={init,next,bad}]
initialize QUBO model $Q$ with $\mathcal{O}(m+|P|)$ init lines in BTOR2 model $B$

// for all $n$ state transitions:
for ($i$ = 1; $i$ <= $n$; $i$++) {

  // updating openat fd bump pointer
  nid_openat_next next 2 nid_fd_bump nid_openat_newfd in BTOR2 model $B$:
    bit-blast nid_openat_newfd into $\mathcal{O}(w)$ BQ function

  // updating brk bump pointer
  nid_brk_next next 2 nid_brk_bump nid_brk_newbrk in BTOR2 model $B$:
    bit-blast nid_brk_newbrk into $\mathcal{O}(w)$ BQ function

  // updating kernel-mode flag
  nid_update_kernel next 1 60 $nid\_kernel\_mode\_flow$ in BTOR2 model $B$:
    bit-blast $nid\_kernel\_mode\_flow$ into $\mathcal{O}(w)$ BQ function

  // updating control flow
  // the total size of the pc-flag BQ functions is still only: $\mathcal{O}(w\cdot|P|)$
  for ($pc$ = $start\_of\_code\_segment$; $pc$ < $end\_of\_code\_segment$; $pc$ = $pc$ + 4)
    nid_$pc$_next next 1 nid_$pc$_flag $nid\_control\_flow$ in BTOR2 model $B$:
      bit-blast $nid\_control\_flow$ into $\mathcal{O}(w\cdot|P|)$ BQ function
    nid_$pc$_stay next 1 nid_$pc$_kernel nid_$pc$_active in BTOR2 model $B$:
      bit-blast nid_$pc$_active into $\mathcal{O}(w\cdot|P|)$ BQ function

  // updating registers
  for ($r$ = 1; $r$ < 32; $r$++)
    nid_$r$_register_update next 2 nid_$r$_register $nid\_register\_flow[r]$ in BTOR2 model $B$:
      bit-blast $nid\_register\_flow[r]$ into $\mathcal{O}(m\cdot w^2\cdot|P|)$ BQ function

  // updating physical memory
  for ($p$ = 0; $p$ < $m$; $p$ = $p$ + 1)
    nid_$p$_memory_update next 2 nid_$p$_RAM_word $nid\_RAM\_write\_flow[p]$ in BTOR2 model $B$:
      bit-blast $nid\_RAM\_write\_flow[p]$ into $\mathcal{O}(w^2\cdot|P|)$ BQ function

  // checking error states
  for ($b$ = 0; $b$ < 12; $b$++)
    nid_check_* bad $nid\_condition$ b$b$ in BTOR2 model $B$:
      bit-blast $nid\_condition$ into $\mathcal{O}(w\cdot|P|)$ BQ function

  compose $\mathcal{O}(m\cdot w^2\cdot|P|)$ BQ function
}

compose $\mathcal{O}(n\cdot m\cdot w^2\cdot|P|)$ QUBO model $Q$
\end{lstlisting}
\caption{QUBOT algorithm}
\label{fig:qubot-algorithm}
\end{figure}

\begin{figure}
  \centering
  \begin{lstlisting}[mathescape,morekeywords={init,next,bad}]
  initialize quantum circuit $C$ with $\mathcal{O}(m+|P|)$ init lines in BTOR2 model $B$

  // for all $n$ state transitions:
  for ($i$ = 1; $i$ <= $n$; $i$++) {

    // updating openat fd bump pointer
    nid_openat_next next 2 nid_fd_bump nid_openat_newfd in BTOR2 model $B$:
      bit-blast nid_openat_newfd into $\mathcal{O}(w)$ quantum circuit

    // updating brk bump pointer
    nid_brk_next next 2 nid_brk_bump nid_brk_newbrk in BTOR2 model $B$:
      bit-blast nid_brk_newbrk into $\mathcal{O}(w)$ quantum circuit

    // updating kernel-mode flag
    nid_update_kernel next 1 60 $nid\_kernel\_mode\_flow$ in BTOR2 model $B$:
      bit-blast $nid\_kernel\_mode\_flow$ into $\mathcal{O}(w)$ quantum circuit

    // updating control flow
    // the total size of the pc-flag BQ functions is still only: $\mathcal{O}(w\cdot|P|)$
    for ($pc$ = $start\_of\_code\_segment$; $pc$ < $end\_of\_code\_segment$; $pc$ = $pc$ + 4)
      nid_$pc$_next next 1 nid_$pc$_flag $nid\_control\_flow$ in BTOR2 model $B$:
        bit-blast $nid\_control\_flow$ into $\mathcal{O}(w\cdot|P|)$ quantum circuit
      nid_$pc$_stay next 1 nid_$pc$_kernel nid_$pc$_active in BTOR2 model $B$:
        bit-blast nid_$pc$_active into $\mathcal{O}(w\cdot|P|)$ quantum circuit

    // updating registers
    for ($r$ = 1; $r$ < 32; $r$++)
      nid_$r$_register_update next 2 nid_$r$_register $nid\_register\_flow[r]$ in BTOR2 model $B$:
        bit-blast $nid\_register\_flow[r]$ into $\mathcal{O}(m\cdot w\cdot|P|)$ quantum circuit

    // updating physical memory
    for ($p$ = 0; $p$ < $m$; $p$ = $p$ + 1)
      nid_$p$_memory_update next 2 nid_$p$_RAM_word $nid\_RAM\_write\_flow[p]$ in BTOR2 model $B$:
        bit-blast $nid\_RAM\_write\_flow[p]$ into $\mathcal{O}(w\cdot|P|)$ quantum circuit

    // checking error states
    for ($b$ = 0; $b$ < 12; $b$++)
      nid_check_* bad $nid\_condition$ b$b$ in BTOR2 model $B$:
        bit-blast $nid\_condition$ into $\mathcal{O}(w\cdot|P|)$ quantum circuit

    compose $\mathcal{O}(m\cdot w\cdot|P|)$ quantum circuit
  }

  compose $\mathcal{O}(n\cdot m\cdot w\cdot|P|)$ quantum circuit $C$
  \end{lstlisting}
  \caption{QUARC algorithm}
  \label{fig:quarc-algorithm}
  \end{figure}

\end{document}